\documentclass[12pt]{article}
\usepackage{amsfonts}
\usepackage[dvips]{epsfig}
\usepackage{graphicx}
\begin{document}

\title{Dynamical Systems, Topology and Conductivity in
Normal Metals.}

\author{A.Ya.Maltsev$^{1}$, S.P.Novikov$^{1,2}$.}

\date{
\centerline{$^{(1)}$ L.D.Landau Institute for Theoretical Physics,}
\centerline{119334 ul. Kosygina 2, Moscow, }
\centerline{ maltsev@itp.ac.ru \,\, ,
\,\,\, novikov@itp.ac.ru}
\centerline{$^{(2)}$ IPST, University of Maryland,}
\centerline{College Park MD 20742-2431,USA}
\centerline{novikov@ipst.umd.edu}}

\maketitle

\begin{abstract}
 New observable integer-valued numbers of the topological
origin were revealed by the present authors studying the
conductivity theory of single crystal 3D normal metals in the
reasonably strong magnetic field ($B \leq 10^{3} Tl$). Our 
investigation is based on the study of dynamical systems on
Fermi surfaces for the motion of semi-classical electron
in magnetic field. All possible asymptotic regimes are also
found for $B \rightarrow \infty$ based on the topological 
classification of trajectories.
\end{abstract}

\section{Introduction.}

This is the survey article dedicated to the topological
approach to the conductivity phenomena in metals in the presence
of the rather strong magnetic field ${\bf B}$. As can be shown,
in the case of rather strong magnetic fields the topology
of the Fermi surface plays the main role in the asymptotic 
behavior of conductivity $\sigma^{ik}$ when $B \rightarrow \infty$.
In this article we represent mainly the results in this area
connected with the "Topological Quantum characteristics" 
introduced recently by the present authors and the classification
results for the regimes of conductivity behavior for
$B \rightarrow \infty$ based on the topological results in
the theory of special dynamical systems.
The main consideration is
made here for the case of the "generic" (or "Topologically
regular") open electron trajectories according to classification
recently introduced in the works of the present authors
(see \cite{novmal1,novmal2, malnov3}). As was shown 
(\cite{novmal1}), this type of 
trajectories always leads to observable topological characteristics 
of the Fermi surface having the form of "stability zones" on the
unit sphere (parameterizing directions of ${\bf B}$) and the
triples of the integer numbers observable in every
"stability zone" in the conductivity measurements. 
Let us say that this type of trajectories corresponds
to the "generic case" and it is the only type of non-closed
trajectories stable with respect to the small perturbations.
Besides that, the trajectories of this type appear with probability 
1 in all cases when the non-closed quasiclassical electron can
be observed on the fixed Fermi surface. 

 However, it can be shown that also other types of open 
orbits corresponding to more complicated "chaotic" behavior can
be observed on the complicated Fermi surfaces. In this case,
the direction of magnetic field should be chosen specially in
the experiment and the behavior of conductivity reveals much
more complicated features. Let us mention also, that the geometric
forms of electron trajectories for such systems can be obtained
as the intersections of the Fermi surface by the planes orthogonal 
to the magnetic field and in some sense these systems are all
"analytically integrable" on the universal covering 
${\mathbb R}^{3}$. However, after the identification of equivalent 
vectors modulo the reciprocal lattice these systems may become
topologically very complicated since the two points in 
${\mathbb R}^{3}$ different by the reciprocal lattice vector
represent actually the same physical state of electron. 
For the generic
irrational directions of ${\bf B}$ the global geometry of these
intersections is apriori unpredictable and can be rather 
complicated as we will see below.

 Let us give here the brief historical survey concerning the
"geometric strong magnetic field limit" in metals. 

 The investigation of the geometric effects in the magnetoresistance 
behavior in normal metals was started first in the school of
I.M.Lifshitz (I.M.Lifshitz, M.Ya.Azbel, M.I.Kaganov, V.G.Peschansky)
in 1950's. Such, in the first paper \cite{lifazkag} the crucial
difference in the asymptotic behavior ($B \rightarrow \infty$)
of conductivity in the cases of closed (compact) 
and open periodic electron
trajectories on the Fermi surface was pointed out. The strong 
anisotropy of conductivity in the last case makes possible the
observation of this phenomenon and the mean direction of the periodic 
open orbits in the experiment. The different examples of the
complicated Fermi surfaces and the corresponding non-closed
electron orbits were considered in the works \cite{lifpes1,lifpes2}.
In particular, the open orbits, more general then just periodic
ones were discovered in \cite{lifpes1}. The geometry of such 
trajectories still was not very complicated and the corresponding
electron motion was still "in average" along the straight line in the
${\bf p}$-space. Again the strong anisotropy of the conductivity
tensor was expected in this case and the form of $\sigma^{ik}$
used in \cite{lifpes1,lifpes2} was actually the same as for
the periodic open orbits. In the paper \cite{lifpes2} the 
important set of analytical dispersion relations was considered 
and the existence of open orbits in the different parts of set 
parameters was discovered. Let us give here also the references
on papers \cite{ag1,ag2,ag3,gaid} (see also the survey articles
\cite{lifkag1,lifkag2} and the book \cite{etm})
where the different question connected with the geometry of open 
orbits for concrete Fermi surfaces of real metals were considered.

 The general problem of classification of different trajectories
arising on arbitrary periodic smooth surface in 3D space as
the intersections with arbitrary planes in ${\mathbb R}^{3}$
(S.P.Novikov problem) was set by S.P.Novikov in \cite{nov1}
and then was considered in his topological school at the
Moscow State University 
(A.V.Zorich, I.A.Dynnikov, S.P.Tsarev, 1980-2000).
\footnote{The general problem of S.P. Novikov has actually more
general form and is connected with the global geometry of level
curves of quasiperiodic functions with $n$ quasiperiods on the
plane. Let us say, however, that the case $n > 3$ was started
to investigate recently (see \cite{nov5}) and is still not so well
studied as the case $n = 3$. Let us make also the reference on the
work \cite{malts2} where the connection of general Novikov problem 
with the specially modulated 2D electron gas was considered.}
Let us mention here that the most important breakthroughs
were made in the papers \cite{zorich,dynn3} where the 
important theorems about the non-compact trajectories were proved.
Using this results the concept of the "Topological quantum numbers"
observable in the conductivity and characterizing the generic
"non-trivial" conductivity behavior in metals was introduced by 
the present authors in \cite{novmal1}. In particular, it was shown 
that these characteristics arise always when the stable
(with respect to the small variations of ${\bf B}$) behavior 
of conductivity different from the "simple" behavior corresponding
to just compact electron orbits on the Fermi surface is observed.
The corresponding "stability zones" on the unit sphere and the
"Triples of integer numbers" ("Topological Quantum characteristics")
give the non-trivial geometric characteristics of the complicated
Fermi surfaces in metals. However, as we already said above
the conductivity can reveal also some "degenerate" (unstable)
behavior in the special cases of rather exotic electron 
trajectories on the Fermi surface. The existence of such 
trajectories as well as the corresponding conductivity behavior 
was studied later in several works 
(\cite{tsarev,dynn4,zorich2,malts}). Let say here that  
the full classification of all types of orbits for arbitrary
smooth periodic surfaces in ${\mathbb R}^{3}$ is finished in
general now (see \cite{dynn4,dynn7}) and the total picture of
different asymptotic behaviors of conductivity (as well as
the conditions for each asymptotic regime) can be described
in main order of $1/B$ in general case (\cite{novmal2,malnov3}).
Let us say also that the general topological investigation 
required rather large set of methods of both 3-dimensional
topology and the dynamical systems theory and we are going just 
to formulate here the corresponding statements needed for our 
purposes.

\section{Some remarks on the Kinetic theory.}

 Let us describe briefly the standard approach to the 
conductivity in crystals
based on one-particle quantum mechanical consideration.
\footnote{Let us note that we use here the Kinetic theory
on the physical level of rigorousness. At the same time all
use of differential topology is based on the mathematically 
rigorous topological results.}
We should consider the Shr\"odinger equation 

$$ - {\hbar^{2} \over 2m} \Delta \psi + V({\bf x}) \psi
= \epsilon \psi $$
in the periodic 
potential $V({\bf x})$ formed by the crystal lattice L: 

$$V({\bf x} + m_{1}{\bf l}_{1} + m_{2}{\bf l}_{2} + 
m_{3}{\bf l}_{3}) \equiv V({\bf x})$$
for all $(m_{1},m_{2},m_{3}) \in {\mathbb Z}^{3}$. 

 The potential $V({\bf x})$
has also all the rotational symmetries of the crystal and in all
physical cases we have the symmetry ${\bf x} \rightarrow -{\bf x}$
after the appropriate choice of the origin in ${\bf x}$-space.

 The reciprocal lattice $\Gamma^{*}$ is generated
by the vectors:

$${\bf g}_{1} = \hbar {{\bf l}_{2} \times {\bf l}_{3} \over
({\bf l}_{1}, {\bf l}_{2}, {\bf l}_{3}) } \,\,\, , \,\,\,
{\bf g}_{2} = \hbar {{\bf l}_{3} \times {\bf l}_{1} \over
({\bf l}_{1}, {\bf l}_{2}, {\bf l}_{3}) } \,\,\, , \,\,\,
{\bf g}_{3} = \hbar {{\bf l}_{1} \times {\bf l}_{2} \over
({\bf l}_{1}, {\bf l}_{2}, {\bf l}_{3}) } $$
in the space of electron momenta.

 The corresponding solutions of the stationary 
Shr\"odinger equation can be represented as Bloch 
functions

$$\psi_{{\bf p},s}({\bf x}) = e^{{i{\bf p}{\bf x} \over \hbar}}
\varphi_{{\bf p},s}({\bf x})$$
where $\varphi_{{\bf p},s}({\bf x})$ has the same periodicity as the
potential $V({\bf x})$.

 The electron states can be parameterized then by the discrete index
$s$ (energy band) and a continuous parameter 
${\bf p} = (p_{1}, p_{2}, p_{3})$ such that the states parameterized 
by ${\bf p}$ and ${\bf p}^{\prime}$ (at the same $s$) different by
any reciprocal lattice vector:

$$n_{1} {\bf g}_{1} + n_{2} {\bf g}_{2} + n_{3} {\bf g}_{3} 
\,\,\,\,\,\, , \,\,\,\,\, (n_{1},n_{2},n_{3}) \in {\mathbb Z}^{3}$$
are physically the same.

 The electron energy $\epsilon$ at fixed $s$ becomes now the function 
of
${\bf p}$, $\epsilon = \epsilon_{s}({\bf p})$, periodic in 
${\bf p}$-space with periods equal to the reciprocal lattice vectors:

$$\epsilon_{s}({\bf p} + 
n_{1} {\bf g}_{1} + n_{2} {\bf g}_{2} + n_{3} {\bf g}_{3}) \equiv
\epsilon_{s}({\bf p}) \,\,\, , \,\,\, 
(n_{1},n_{2},n_{3}) \in {\mathbb Z}^{3} $$

 The full set of functions $\epsilon_{s}({\bf p})$ inherits also
the rotational symmetry of the potential $V({\bf x})$ and for any
function $\epsilon_{s}({\bf p})$ we have a symmetry

$$\epsilon_{s}({\bf p}) = \epsilon_{s}(-{\bf p})$$
after the appropriate choice of the initial point in 
${\bf p}$-space.
 
 From topological point of view the correct phase space for any
fixed energy band is the three-dimensional compact torus
(Brillouin zone) ${\mathbb T}^{3} = {\mathbb R}^{3}/G$ rather then 
the open 
three-dimensional space ${\mathbb R}^{3}$. The functions 
$\epsilon_{s}({\bf p})$ become then the (smooth) functions on
${\mathbb T}^{3}$ with the same property 
$\epsilon_{s}({\bf p}) = \epsilon_{s}(-{\bf p})$ for the
appropriate initial point in ${\mathbb T}^{3}$.

 The collective electron
structure can be described 
by the one-particle distribution functions $f_{s}({\bf p})$
satisfying to Boltzmann equation.

 According to the classical results in the absence of external 
fields the functions $f_{s}({\bf p})$ are given by the
Fermi distribution

\begin{equation}
\label{fermdist}
f_{s}({\bf p}) = 
{1 \over 1 + e^{{\epsilon_{s}({\bf p}) - \epsilon_{F} \over T}}}
\end{equation}
for the fixed temperature $T$.

 Let us mention also that every function $\epsilon_{s}({\bf p})$
is bounded by some minimal and maximal values $\epsilon_{s}^{min}$,
$\epsilon_{s}^{max}$ being the smooth function on the compact 
manifold.

 For the case of normal metals we will always have a situation
$T \ll \epsilon_{F}$ and the distribution function will change only
in narrow energy interval $(\sim T)$ near the Fermi energy
$\epsilon_{s}({\bf p}) \sim \epsilon_{F}$, being equal to $1$
for $\epsilon_{F} - \epsilon_{s}({\bf p}) \gg T$ and to $0$
for $\epsilon_{s}({\bf p}) - \epsilon_{F} \gg T$. So, the energy 
bands with $\epsilon_{s}^{max} < \epsilon_{F}$ will be completely
filled in this case ($f_{s}({\bf p}) \equiv 1$) and the energy
bands with $\epsilon_{s}^{min} > \epsilon_{F}$ will be completely
empty ($f_{s}({\bf p}) \equiv 0$).

 Geometrically we have occupied all the states "inside"
the surface $\epsilon_{s}({\bf p}) = \epsilon_{F}$ 
in 3D torus and empty all
the states "outside". The union of surfaces 
$\epsilon_{s}({\bf p}) = \epsilon_{F}$ for all $s$ such that
$\epsilon_{s}^{min} < \epsilon_{F} < \epsilon_{s}^{max}$ gives then
the total Fermi surface $S_{F}$ of the given metal.

 The Fermi surface will be called non-singular if 
$\nabla \epsilon_{s}({\bf p}) \neq 0$ for all $s$ on the Fermi 
level $\epsilon_{F}$, otherwise we will call it singular.

 It will be convenient to consider also the natural covering
${\hat S}_{F}$ over the Fermi surface in ${\mathbb R}^{3}$ which is 
a smooth 3-periodic surface (for non-singular $S_{F}$) in the
${\bf p}$-space with the periods equal to the reciprocal lattice
vectors.

 The gradient of function $\epsilon_{s}({\bf p})$ is the group
velocity of a particle in the corresponding quantum state and 
describes the mean velocity of the localized wave packet formed
by the wave functions $\psi_{s,{\bf p}^{\prime}}({\bf x})$ with
values ${\bf p}^{\prime}$ close to ${\bf p}$. Easy to see that the 
group velocity is the odd function of ${\bf p}$ (after the
appropriate choice of the initial point)

$${\bf v}_{s\,\, gr} (-{\bf p}) = - {\bf v}_{s\,\, gr}({\bf p})$$
since all the functions $\epsilon_{s}({\bf p})$ are even. The total 
flux density

$${\bf i} = 2 \sum_{s} \int\dots\int 
{\bf v}_{s\,\, gr}({\bf p}) \, f_{s}({\bf p}) 
{d^{3} p \over (2\pi\hbar)^{3}} = 
2 \sum_{s} \int\dots\int
\nabla \epsilon_{s}({\bf p}) \, f_{s}({\bf p})
{d^{3} p \over (2\pi\hbar)^{3}}$$
as well as the electric currents density ${\bf j} = e {\bf i}$
are zero for the equilibrium functions given by (\ref{fermdist}).
In our further consideration we will deal with the small deviations
from the Fermi distributions (\ref{fermdist}) caused by small
external forces (electric field) and the correspondent small
response in the total electric current density.

 Let us put now 
${\bf v}_{gr}({\bf p}) = {\bf v}_{s\,\, gr}({\bf p})$ on the 
corresponding components of the Fermi surface and consider just one
function ${\bf v}_{gr}({\bf p})$ on $S_{F}$. 

 We will use now the fact that all the components
of $S_{F}$ give the independent contributions to conductivity tensor
$\sigma^{ik}$. This means that  we can actually consider only one 
conductivity zone with
the dispersion relation $\epsilon({\bf p}) = \epsilon_{s}({\bf p})$
and the corresponding distribution function 
$f({\bf p}) = f_{s}({\bf p})$ defined everywhere in 
${\mathbb T}^{3}$. The 
corresponding contributions to the electric conductivity (in the
presence of magnetic field) should be then just added in the
3-dimensional tensor $\sigma^{ik}$.

 So we will consider now only one dispersion relation 
$\epsilon({\bf p})$ and the function $f({\bf p})$ which satisfies
to the Boltzmann equation for the spatially homogeneous case
(we omit here the spin dependence of the distribution function
for simplicity):

\begin{equation}
\label{boltzeq1}
{\partial f \over \partial t} + 
{\bf F}^{ext} {\partial f \over \partial {\bf p}} = 
I[f]({\bf p})
\end{equation}

 Here ${\bf F}^{ext}$ is the homogeneous external force and
$I[f]({\bf p})$ is the collision integral. For our situation of low
temperatures ($T \sim 1K$) only the scattering on the impurities
will play the main role so we will write here the functional
$I[f]({\bf p})$ in the general form:

$$I[f]({\bf p}) = \int\dots\int \left[ 
f({\bf p}^{\prime})(1 - f({\bf p})) - 
f({\bf p})(1 - f({\bf p}^{\prime})) \right] 
\sigma ({\bf p},{\bf p}^{\prime}) 
{d^{3} p^{\prime} \over (2\pi\hbar)^{3}} = $$

$$= \int\dots\int \left[
f({\bf p}^{\prime}) - f({\bf p}) \right]
\sigma ({\bf p},{\bf p}^{\prime})
{d^{3} p^{\prime} \over (2\pi\hbar)^{3}} $$

 Using of the Boltzmann equation implies the 
applicability of the semi-classical approach for the electron motion 
and the evolution of the quantum state within the Brillouin zone can
be described just by the "classical" dynamical system:

$${\dot {\bf p}} = {\bf F}^{ext}$$
with the only difference that ${\bf p}$ now is a quasimomentum
and belongs to 3-dimensional torus 
${\mathbb T}^{3}$ instead of the Euclidean 
space. 

 In our case we will have

$${\bf F}^{ext} = {e \over c} 
\left[{\bf v}_{gr} \times {\bf B} \right] + e {\bf E} =
\left[\nabla \epsilon({\bf p}) \times {\bf B} \right] + e {\bf E} $$
where ${\bf B}$ is the homogeneous magnetic field and ${\bf E}$
is the small electric field responsible for the electric current.

 The applicability of semiclassical approach requires then that
the fields ${\bf B}$ and ${\bf E}$ are small with respect to the
internal fields in the crystal. We will require also the condition
$\hbar \omega_{B} \ll \epsilon_{F}$ where $\omega_{B} = eB/m^{*}c$ 
is the cyclotron frequency which will permit to consider the
quantization of levels as a small effect with respect to the
classical motion picture. The electric field is going to be
infinitesimally small so we don't put any conditions on it.
Let us just mention here that these semiclassical conditions
are satisfied very well for normal metals in all experimentally
available fields ${\bf B}$ (the theoretical limit is 
$B \sim 10^{3}-10^{4} Tl$).
Thus we use the Boltzmann equation (\ref{boltzeq1}) to describe
the main part of the conductivity tensor $\sigma^{ik}$ while
the quantum phenomena will be considered as the small corrections
in our situation.

 Now the only reminiscent of the quantization will be the 
dispersion relation $\epsilon({\bf p})$ and the corresponding
dynamical system on ${\mathbb T}^{3}$ given by

$${\dot {\bf p}} = {e \over c}
\left[\nabla \epsilon({\bf p}) \times {\bf B} \right] + e {\bf E} $$

 For the stationary distribution function $f({\bf p})$ we can write
now the Boltzmann equation as

\begin{equation}
\label{boltzeq2}
{e \over c} \left[\nabla \epsilon({\bf p}) \times {\bf B} \right]
\nabla f({\bf p}) + e {\bf E} \nabla f({\bf p}) =
I[f]({\bf p})
\end{equation}

 Any Fermi distribution function (\ref{fermdist}) satisfies the
condition $I[f_{0}]({\bf p}) \equiv 0$ and gives an equilibrium
distribution for given temperature $T$ in the absence of external
fields. 

 Let us now consider in details the main dynamical system:

\begin{equation}
\label{dynsyst}
{\dot {\bf p}} = {e \over c}
\left[\nabla \epsilon({\bf p}) \times {\bf B} \right]
\end{equation}
on ${\mathbb T}^{3}$ for general periodic $\epsilon({\bf p})$.

 System (\ref{dynsyst}) is Hamiltonian with respect to the
(non-standard) Poisson bracket

\begin{equation}
\label{pb}
\{p_{1},p_{2}\} = {e \over c} B_{3} \,\,\, , \,\,\,
\{p_{2},p_{3}\} = {e \over c} B_{1} \,\,\, , \,\,\,
\{p_{3},p_{1}\} = {e \over c} B_{2}
\end{equation}
with the Hamiltonian function  $\epsilon({\bf p})$. The bracket
(\ref{pb}) is degenerate with the Casimir

$$C = \sum_{i=1}^{3} p_{i} B_{i} = ({\bf p} \cdot {\bf B}) $$

 However, the Casimir function $C$ is a multi-valued function
on the three-dimensional torus ${\mathbb T}^{3}$ and should be actually
considered as a $1$-form in it. Geometrically, the trajectories 
of (\ref{dynsyst}) in ${\mathbb T}^{3}$ will be given on every energy
level $\epsilon({\bf p}) = const$ by the level curves of this
$1$-form restricted on these two-dimensional surfaces. So we 
have locally the analytic integrability of the system 
(\ref{dynsyst}). However, the global geometry of the trajectories
of (\ref{dynsyst}) can be highly nontrivial because of the
non-uniqueness of values of Casimir function $C$. 

 The picture becomes more visible if we consider the corresponding
covering 
${\mathbb R}^{3} \rightarrow {\mathbb T}^{3}$ 
defined by the reciprocal
lattice in ${\bf p}$-space. The corresponding function
$\epsilon({\bf p})$ is then the three-periodic function in 
${\mathbb R}^{3}$.
The Fermi surface (the covering over the Fermi surface in 
${\mathbb T}^{3}$)
also becomes the three-periodic (smooth) surface in the 
${\bf p}$-space. The Casimir function $C$ is now well defined
function in ${\mathbb R}^{3}$ (height function) and the trajectories 
of (\ref{dynsyst}) on the Fermi level will be given by the 
intersections of the three-periodic surface 
$\epsilon({\bf p}) = \epsilon_{F}$ with all different planes
orthogonal to ${\bf B}$.
The most important thing for us will be the global geometry of
the trajectories in the planes $\Pi({\bf B})$ orthogonal to the
magnetic field. 

 The dynamical system (\ref{dynsyst}) conserves also the volume 
element $d^{3} p$ in 
${\mathbb T}^{3}$ and does not change at all the
Fermi distribution (\ref{fermdist}). So, in the absence of the
electric field ${\bf E}$ we will have the electron distribution
unchanged (up to the quantum corrections). Nevertheless, the 
response of this system to small perturbations will be completely
different from the case ${\bf B} = 0$ and depend strongly on the 
geometry of trajectories of the dynamical system (\ref{dynsyst}).

 Let us now come back to the Boltzmann equation (\ref{boltzeq2})
and analyze the effects of infinitesimally small electric field
${\bf E}$. 

 We will use for simplicity the $\tau$-approximation for the 
collision integral
throughout this paper and just put 
for small perturbations $\delta f$:

$$I\left[f_{T} + \delta f \right]({\bf p}) - 
I\left[f_{T} \right]({\bf p}) =
{\hat J}_{T} \left( \delta f \right) ({\bf p}) = 
- {1 \over \tau} \delta f ({\bf p}) $$

 Let us make now also some (standard) remarks about $\tau$ for 
our case of normal metals at low temperatures ($ \sim 1K$). 
As it is well-known for the low temperatures both the 
electron-phonon and electron-electron scattering decrease 
(as $T^{5}$ and $T^{2}$ respectively) and the only reminiscent
at $T \rightarrow 0$ is the scattering by the impurities
which is constant for $T \rightarrow 0$. For the elastic scattering
by impurities we can usually write the collision integral in the
form

$$I\left[f \right]({\bf p}) = {2\pi \over \hbar} n_{i}
\int\dots\int |v_{{\bf p}{\bf p^{\prime}}}|^{2}
\left[f({\bf p^{\prime}}) - f({\bf p})\right]
\delta (\epsilon({\bf p}) - \epsilon({\bf p^{\prime}})
{d^{3} p \over (2\pi\hbar)^{3}} $$
where $n_{i}$ is the concentration of impurities and 
$v_{{\bf p}{\bf p^{\prime}}}$ is the amplitude of scattering
${\bf p} \rightarrow {\bf p^{\prime}}$ on the impurity.
Such we can put
$$I\left[f \right]({\bf p}) =
{2\pi \over \hbar} n_{i} \int\int_{S_{F}}
|v_{{\bf p}{\bf p^{\prime}}}|^{2}
\left[f({\bf p^{\prime}}) - f({\bf p})\right]
{dS^{\prime} \over v_{gr}({\bf p^{\prime}})} $$
on the Fermi surface.

 Easy to see then that we have the property $I[f] \equiv 0$
for any function $f$ depending only on energy: 
$f({\bf p}) = f(\epsilon({\bf p}))$. So for this form of $I[f]$
the corresponding operator ${\hat J}_{T}$ will have a lot of
zero modes corresponding to the perturbations
$\delta f({\bf p}) = \delta f(\epsilon({\bf p}))$. Practically
this means that the processes of electron-electron and
electron-phonon scattering responsible for the mixing of
different energy levels become very small at low temperatures.
The corresponding "energy relaxation time" 
$\tau_{en} \rightarrow \infty$ as $T \rightarrow 0$ and the
characteristic times of energy relaxation become very big at low
$T$. 

 However the energy relaxation time $\tau_{en}$ will not be
important for us in the consideration of the conductivity
phenomena. Indeed, the mean electric flux density given by

$${\bf j} = e \int\dots\int {\bf v}_{gr}({\bf p})
\delta f ({\bf p}) {d^{3} p \over (2\pi\hbar)^{3}} $$
remains zero for any distribution perturbation of form
$\delta f({\bf p}) = \delta f(\epsilon({\bf p}))$ because of
the property ${\bf v}_{gr}(-{\bf p}) = - {\bf v}_{gr}({\bf p})$.
We have so that all the zero modes of ${\hat J}_{T}$ do not
give any contribution to the conductivity and we can 
omit them in our consideration.

 We will consider now only the perturbations $\delta f({\bf p})$
such that the total number of particles on each energy level 
remains unchanged w.r.t. the initial Fermi distribution. So we
put an additional restriction

\begin{equation}
\label{restr}
\int\int_{\epsilon({\bf p}) = const} \delta f({\bf p})
{d S \over v_{gr}({\bf p})} \equiv 0
\end{equation}
for all the energy levels. Let us say also that the only
factor which can disturb this condition according to the full
Boltzmann equation

$${\partial f \over \partial t} + {e \over c} \left[
\nabla \epsilon({\bf p}) \times {\bf B} \right]
{\partial f \over \partial {\bf p}} + 
e {\bf E} {\partial f \over \partial {\bf p}} =
I \left[ f \right] $$
is the electric field ${\bf E}$. However, because of the
equality

$$\int\dots\int {\bf E} 
{\partial f_{T} \over \partial {\bf p}}
{d^{3} p \over (2\pi\hbar)^{3}} = 
\int\dots\int {\bf E} \nabla \epsilon({\bf p})
{\partial f_{T} \over \partial \epsilon}
{d^{3} p \over (2\pi\hbar)^{3}} \equiv 0$$
this effect has the order $E^{2}$ and should be omitted in the
computing a linear approximation for the infinitesimally small
$E$. We can now forget about the zero modes of the functional
${\hat J}_{T}$ using the functional space defined by (\ref{restr})
and put $\tau$ to be the characteristic relaxation time on each
energy level due to the scattering on the impurities.
This time $\tau = \tau_{mom}$ is then the momentum relaxation
time and it is this quantity which is responsible for
conductivity at $T \rightarrow 0$. We just use then the
approximation

$${\hat J}_{T} \left[ \delta f \right]({\bf p}) = 
- {1 \over \tau} \delta f ({\bf p})$$
on the space of function satisfying (\ref{restr}) where
$\tau = \tau_{mom}$ has the meaning of the free electron 
motion time
before the scattering by the impurity and remains constant at
$T \rightarrow 0$.

 The Boltzmann equation (\ref{boltzeq2}) can be rewritten now
in linear order of $E$ as

\begin{equation}
\label{tauappr}
{e \over c} \left[\nabla \epsilon({\bf p}) \times {\bf B} \right]
\nabla f_{1}({\bf p}) + e {\bf E} {\bf v}_{gr}({\bf p})
{\partial f_{T} \over \partial \epsilon} = 
- {1 \over \tau} f_{1} ({\bf p})
\end{equation}
where $f_{1}({\bf p})$ is a linear in $E$ perturbation of
$f_{T}$.

 Let us introduce the vector field ${\bf \xi}({\bf p})$ on
${\mathbb T}^{3}$ according to our system (\ref{dynsyst}):

$${\bf \xi}({\bf p}) = 
{e \over c} \left[\nabla \epsilon({\bf p}) \times {\bf B} 
\right] $$

 We can then rewrite the equation (\ref{tauappr}) on $f_{1}$
in the form:

\begin{equation}
\label{tap2}
\nabla_{\bf \xi} f_{1}({\bf p}) + {1 \over \tau} f_{1} ({\bf p})
= - e {\bf E} {\bf v}_{gr}({\bf p})
{\partial f_{T} \over \partial \epsilon}
\end{equation}

 The equation (\ref{tap2}) can be now solved separately on every
trajectory of the system (\ref{dynsyst}). Indeed, let us 
introduce the coordinate $t$ (time) along the trajectories
of (\ref{dynsyst}). We have then

$$e^{-t/\tau} {d \over dt} \left[ e^{t/\tau} f_{1}(t) \right] =
e \left(-{\partial f_{T} \over \partial \epsilon}\right)
{\bf E} {\bf v}_{gr}(t) $$
on every trajectory. We put here $f_{1} = f_{1}(t)$,
${\bf v}_{gr} = {\bf v}_{gr}(t)$ and the value 
$\partial f_{T}/\partial \epsilon$ is constant on the trajectories
of (\ref{dynsyst}).

 The solution of (\ref{tap2}) can be written now as (\cite{etm}):

$$f_{1}(t) = 
e \left(-{\partial f_{T} \over \partial \epsilon}\right)
\int_{-\infty}^{t} e^{{t^{\prime}-t \over \tau}}
{\bf E} {\bf v}_{gr}(t^{\prime}) dt^{\prime} =$$
\begin{equation}
\label{solution}
= e \left(-{\partial f_{T} \over \partial \epsilon}\right)
\int_{0}^{+\infty} e^{-t^{\prime} \over \tau} {\bf E} 
{\bf v}_{gr}(t-t^{\prime}) dt^{\prime}
\end{equation}

 Let us introduce the "geometric" time $s = teB/c$ where $t$ is
the time along the trajectories according to the system
(\ref{dynsyst}). For the derivative of ${\bf p}$ with respect to
$s$ we then will have the system:

\begin{equation}
\label{sder}
{d {\bf p} \over ds} = \left[ \nabla \epsilon({\bf p}) \times
{\bf n} \right]
\end{equation}
where ${\bf n} = {\bf B}/B$ is a unit vector along ${\bf B}$.
Let us introduce the notation 
${\bf v}_{gr}({\bf p},t) = {\bf v}_{gr}({\bf p}(t))$
where ${\bf p}(t)$ is a solution of the system (\ref{dynsyst})
with the initial data ${\bf p}(0) = {\bf p}$. The formula
(\ref{solution}) can be then written as

\begin{equation}
\label{sol2}
f_{1}({\bf p}) = 
e \left(-{\partial f_{T} \over \partial \epsilon}\right)
\int_{0}^{+\infty} 
{\bf E} {\bf v}_{gr}({\bf p},-t)
e^{-t/\tau} dt 
\end{equation}

 The parameter $s$ plays the role of geometric parameter
along the trajectories, such that $dl/ds = v^{\perp}_{gr}(s)$
where $dl$ is the length element in the ${\bf p}$-space
and $v^{\perp}_{gr}(s)$ is the length of component of 
${\bf v}_{gr}(s)$ orthogonal to ${\bf B}$.
The size of the Brillouin zone corresponds then to
$s \sim p_{F}/v_{F} \sim m^{*}$ where $m^{*}$ is some parameter
with mass dimension playing the role of "typical" electron mass
on the Fermi surface for a given dispersion relation.

 We will be interested  in the "strong geometric limit" when the
parameter $m^{*}c/eB\tau$ is much less then $1$. This
condition can be also written as $\omega_{B}\tau \gg 1$ where
$\omega_{B} = {eB/m^{*}c}$ is the cyclotron frequency corresponding
to the mass $m^{*}$. It is easy to see from (\ref{sol2}) that this
case corresponds to "long integrations" of ${\bf v}_{gr}({\bf p})$
along the trajectories of (\ref{sder}) (much longer then one
Brillouin zone for open trajectories) such that the global geometry
of the trajectories becomes the most important factor for the
linear response $f_{1}({\bf p})$. 

 The correspondent current density ${\bf j}$ has the form

$${\bf j} = e \int\dots\int 
{\bf v}_{gr}({\bf p}) f_{1}({\bf p})
{d^{3} p \over (2\pi\hbar)^{3}} = $$

$$= e^{2} \int\dots\int {\bf v}_{gr}({\bf p})
\left(-{\partial f_{T} \over \partial \epsilon}\right)
\left[\int_{0}^{+\infty}
{\bf E} {\bf v}_{gr}({\bf p},-t) e^{-t/\tau} dt \right]
{d^{3} p \over (2\pi\hbar)^{3}} $$
and we have for the conductivity tensor $\sigma^{ik}$
(\cite{etm}, \S 27):

\begin{equation}
\label{sigmaik}
\sigma^{ik}({\bf B}) = e^{2} \int\dots\int
\left(-{\partial f_{T} \over \partial \epsilon}\right)
v^{i}_{gr}({\bf p}) \left[\int_{0}^{+\infty}
v^{k}_{gr}({\bf p},-t) e^{-t/\tau} dt \right]
{d^{3} p \over (2\pi\hbar)^{3}} 
\end{equation}

 Easy to prove also the Onsager relations for tensor
$\sigma^{ik}({\bf B})$:

$$\sigma^{ik}({\bf B}) = \sigma^{ki}(-{\bf B})$$.

 It is easy to see that the expression 

$$\int\dots\int
\left(-{\partial f_{T} \over \partial \epsilon}\right)
v^{i}_{gr}({\bf p}) v^{k}_{gr}({\bf p},-t)
{d^{3} p \over (2\pi\hbar)^{3}} $$
is invariant under the action of the dynamical system 
(\ref{dynsyst}).
We have so:

$$\sigma^{ik}({\bf B}) = e^{2} \int_{0}^{+\infty}
\left[ \int\dots\int
\left(-{\partial f_{T} \over \partial \epsilon}\right)
v^{i}_{gr}({\bf p}) v^{k}_{gr}({\bf p},-t)
{d^{3} p \over (2\pi\hbar)^{3}} \right]
e^{-t/\tau} dt =$$

\begin{equation}
\label{sigma}
= e^{2} \int_{0}^{+\infty}
\left[ \int\dots\int
\left(-{\partial f_{T} \over \partial \epsilon}\right)
v^{i}_{gr}({\bf p},t) v^{k}_{gr}({\bf p})
{d^{3} p \over (2\pi\hbar)^{3}} \right]
e^{-t/\tau} dt 
\end{equation}

 Using now the fact that $v^{i}_{gr}({\bf p},t)$ coincides
with the value $v^{i}_{gr}({\bf p},-t)$ defined for the system
(\ref{dynsyst}) for $-{\bf B}$ we get the Onsager identity.

\section{Topological considerations and theorems.}

 Before starting the investigation of the dependence of 
${\sigma}^{ik}$ on the geometry of orbits of (\ref{dynsyst})
let us make here some basic 
definitions and formulate the topological results which we are 
going to use in our considerations. 

\begin{figure}
\begin{center}
\epsfig{file=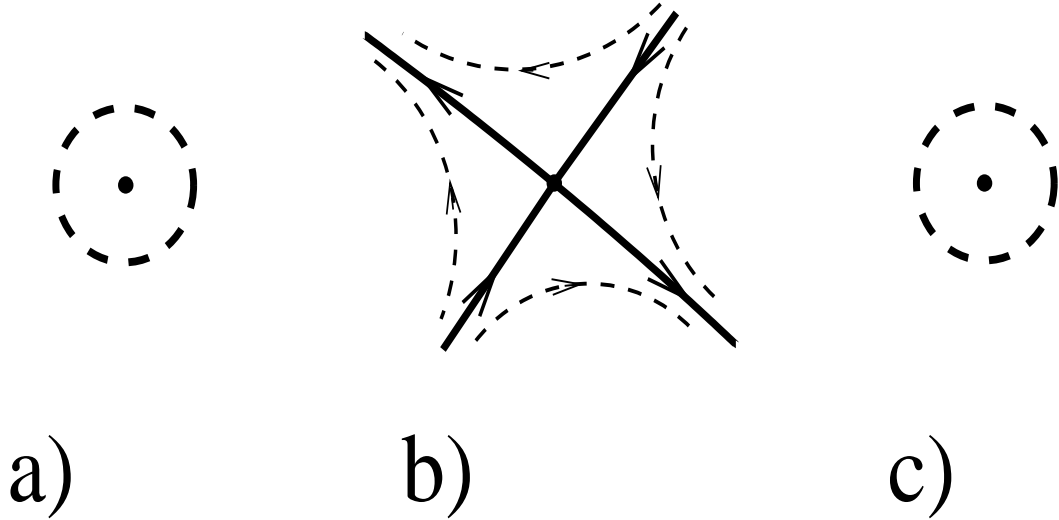,width=14.0cm,height=7cm}
\end{center}
\caption{The level curves of the function
$\epsilon({\bf p})|_{\Pi}$ in the plane close to the local minimum,
the saddle-point and the local maximum of
$\epsilon({\bf p})|_{\Pi}$.
}
\label{critpoints}
\end{figure}

 First of all, we consider the intersections of the smooth 
3-periodic Fermi surface in ${\bf p}$-space, with the set of
parallel planes orthogonal to magnetic field ${\bf B}$. We
can consider the corresponding trajectories also as the level 
curves of the restrictions $\epsilon({\bf p})|_{\Pi}$ of the
dispersion relation to the planes orthogonal to ${\bf B}$
which will then be the quasiperiodic functions in this planes
with 3 quasiperiods according to the standard definition.
Let us make now the generic assumption that the restrictions 
$\epsilon({\bf p})|_{\Pi}$ are the Morse functions in any
plane $\Pi$ orthogonal to ${\bf B}$, i.e. all the critical
points of this functions can be the non-degenerate local minima,
non-degenerate saddle points or the non-degenerate local
maxima in the plane. On the Fig. \ref{critpoints} we show the 
local behavior
of the trajectories close to these critical points and we
don't admit here any other degeneration of trajectories in the 
plane.

Let us give here the following definitions:

\vspace{0.5cm}

 {\bf Definition 1.} {\it We call the trajectory non-singular if it
is not adjacent to the critical (saddle) point of the function
$V({\bf r})$. The trajectories adjacent to the critical points
as well as the critical points themselves we call singular
trajectories (see Fig. \ref{critpoints}).}

\vspace{0.5cm}

{\bf Definition 2.} {\it We call the non-singular trajectory compact
if it is closed on the plane. We call the non-singular trajectory
open if it is unbounded in ${\mathbb R}^{2}$.}
 
\vspace{0.5cm}

\begin{figure}
\begin{center}
\epsfig{file=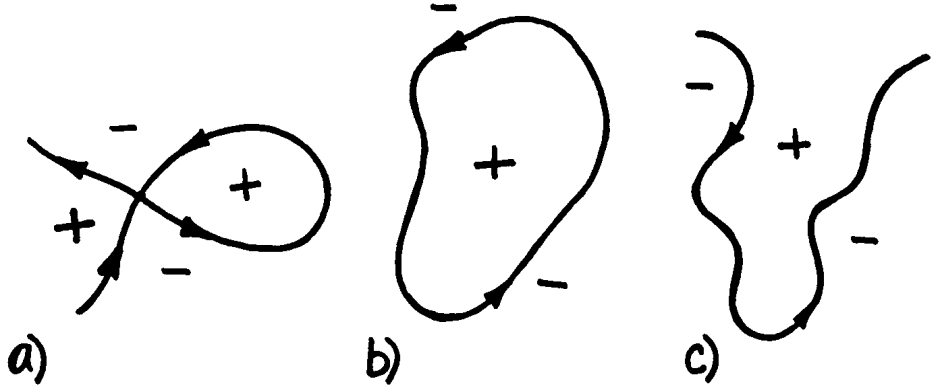,width=14.0cm,height=8cm}
\end{center}
\caption{The singular, compact and open non-singular
quasiclassical trajectories. The signs $"+"$ and $"-"$ show the  
regions of larger and smaller values of
$\epsilon({\bf p})|_{\Pi}$ respectively.
}
\label{traject}
\end{figure}

 The examples of singular, compact and open non-singular
trajectories are shown on the Fig. \ref{traject}, a-c.

 It is easy to see also that the singular trajectories have the
measure zero among all the trajectories on the plane.

\begin{figure}
\begin{center}
\epsfig{file=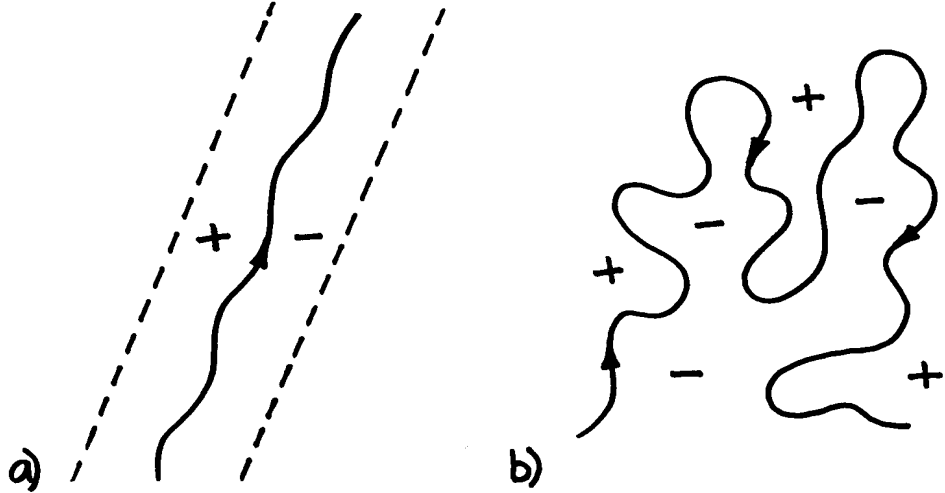,width=14.0cm,height=8cm}
\end{center}
\caption{"Topologically regular" (a) and "chaotic"
(b) open trajectories in the plane $\Pi$ orthogonal to ${\bf B}$.
}
\label{regandch}
\end{figure}

 In our consideration we will not make any difference between the 
compact trajectories of different forms and consider the global
geometry of the non-compact trajectories only. According to the
topological results on the classification of non-compact orbits
it will be convenient to use the following definition concerning
the global behavior of open trajectories in the planes orthogonal
to ${\bf B}$:

\vspace{0.5cm}

 {\bf Definition 3.} {\it We call the open trajectory topologically
regular (corresponding to "topologically integrable" case)
if it lies within the straight line of finite width
in ${\mathbb R}^{2}$ and passes through it from $-\infty$ to   
$\infty$ (see Fig. \ref{regandch}, a). All other open trajectories 
we will call chaotic (Fig. \ref{regandch}, b).}
 
\vspace{0.5cm}

 We will see below that this approach to the classification
of electron trajectories is closely connected with the form
of the Fermi surface itself and the forms of the "carriers
of non-compact trajectories" belonging to the Fermi surface.
Thus the "topological integrability" is actually the property
characterizing the topology of the carriers 
of open trajectories in ${\mathbb T}^{3}$ (and 
${\mathbb R}^{3}$) and has the direct analog in the theory of 
integrable systems although the reasons of "integrability"
are completely different in this situation and have completely
topological origin. 

 Let us give now the basic definitions concerning the topology
of the Fermi surface and later the topology of 
"carriers of open trajectories" on the Fermi surface
for the fixed direction of ${\bf B}$.

\vspace{0.5cm}

{\bf Definition 4.} 

\begin{figure}
\begin{center}
\epsfig{file=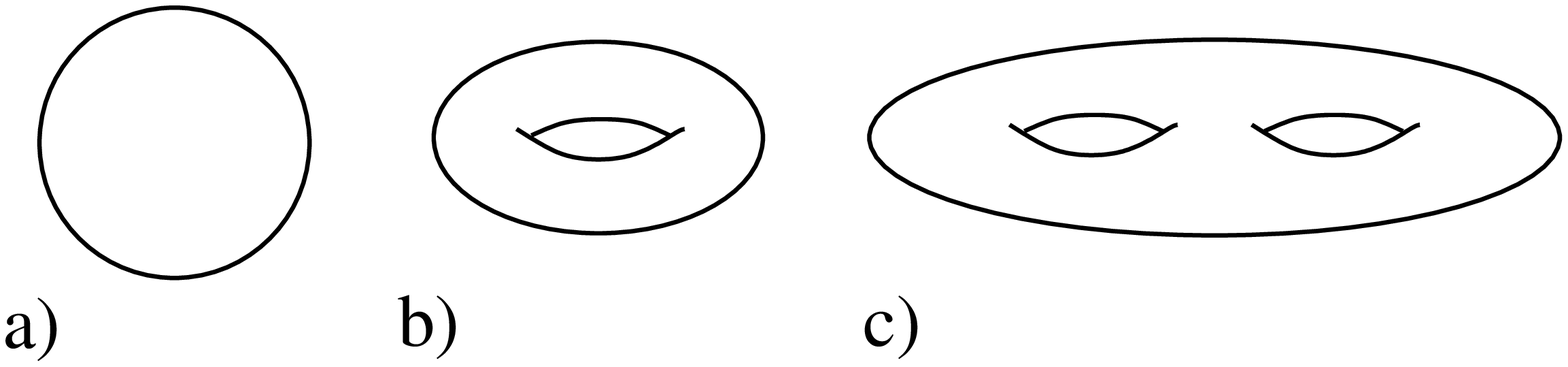,width=14.0cm,height=3.5cm}
\end{center}
\caption{The abstract surfaces with genuses $0$, $1$ and $2$
respectively.
}
\label{genus}
\end{figure}

{\it I) Genus.

 Let us now come back to the original phase space
${\mathbb T}^{3} = {\mathbb R}^{3}/\Gamma^{*}$. 
 Every component of the Fermi surface becomes then the smooth    
orientable 2-dimensional surface embedded in ${\mathbb T}^{3}$. 
We can then
introduce the standard genus of every component of the Fermi
surface $g = 0,1,2,...$ according to standard topological
classification depending on if this component is topological
sphere, torus, sphere with two holes, etc ...  
(see Fig. \ref{genus}).

II) Topological Rank.

 Let us introduce the Topological Rank $r$ as the characteristic
of the embedding of the Fermi surface in ${\mathbb T}^{3}$. 
It's much more
convenient in this case to come back to the total ${\bf p}$-space
and consider the connected components of the three-periodic 
surface in ${\mathbb R}^{3}$.

1) The Fermi surface has Rank $0$ if every its connected component
can be bounded by a sphere of finite radius.

2) The Fermi surface has Rank $1$ if every its connected component
can be bounded by the periodic cylinder of finite radius and there
are components which can not be bounded by the sphere.

3) The Fermi surface has Rank $2$ if every its connected component
can be bounded by two parallel (integral) planes in 
${\mathbb R}^{3}$ and  
there are components which can not be bounded by cylinder.  

4) The Fermi surface has Rank $3$ if it contains components which
can not be bounded by two parallel planes in 
${\mathbb R}^{3}$. }

\vspace{0.5cm}

 On the Fig. \ref{genus} we show the topological form of the 
surfaces of genuses 0, 1 and 2 (without the embedding in 
${\mathbb T}^{3}$). It is easy to generalize this picture to
the arbitrary genus $g$.

\begin{figure}
\begin{center}
\epsfig{file=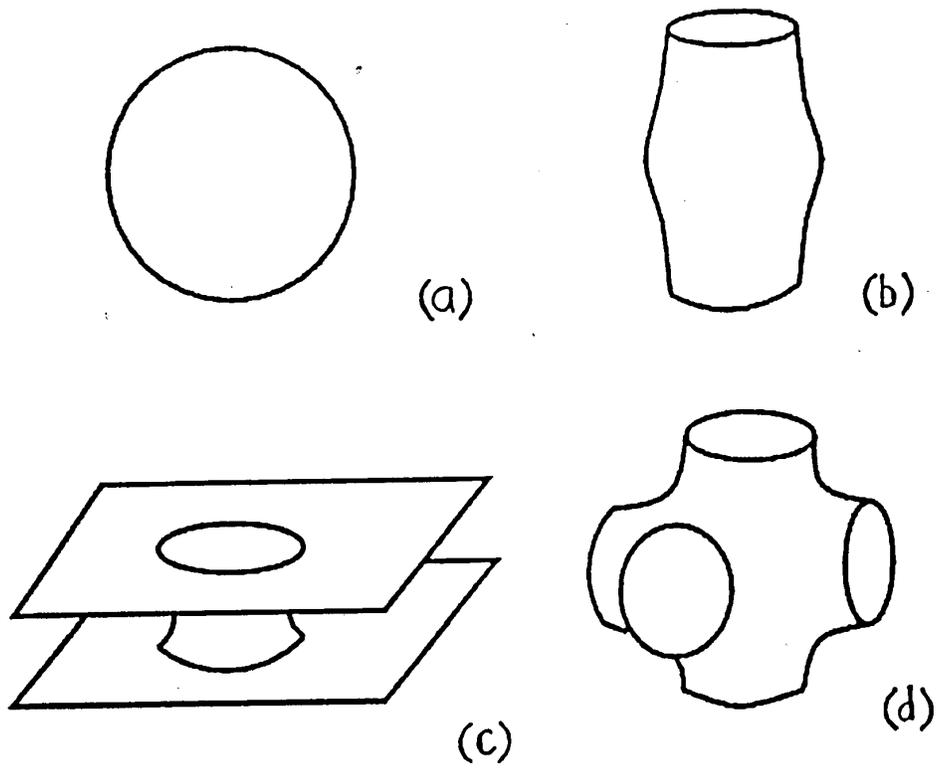,width=14.0cm,height=11cm}
\end{center}
\caption{The Fermi surfaces with Topological Ranks $0$, $1$, $2$
and $3$ respectively.}
\label{rank}
\end{figure}

 The pictures on Fig. \ref{rank}, a-d represent the pieces of 
the Fermi surfaces in ${\mathbb R}^{3}$ with the Topological Ranks 
$0$, $1$, $2$ and $3$ respectively.

 It is easy to see also that the topological Rank coincides with the
maximal Rank of the image of mapping
$\pi_{1}(S^{i}) \rightarrow \pi_{1}({\mathbb T}^{3})$ 
for all the connected components of the Fermi surface.

 As can be seen the genuses of the surfaces
represented on the Fig. \ref{rank}, a-d are also equal to 
$0$, $1$, $2$ and   
$3$ respectively. However, the genus and the Topological Rank are not
necessary equal to each other in the general situation.

 Let us discuss briefly the connection between the genus and the
Topological Rank since this will play the crucial role in further
consideration. It is easy to see that the Topological Rank of the
sphere can be only zero and the Fermi surface consists in this case
of the infinite set of the periodically repeated spheres 
${\mathbb S}^{2}$ in ${\mathbb R}^{3}$.

 The Topological Rank of the torus ${\mathbb T}^{2}$ can take
three values $r = 0$, $r = 1$ and $r = 2$.

\begin{figure}
\begin{center}
\epsfig{file=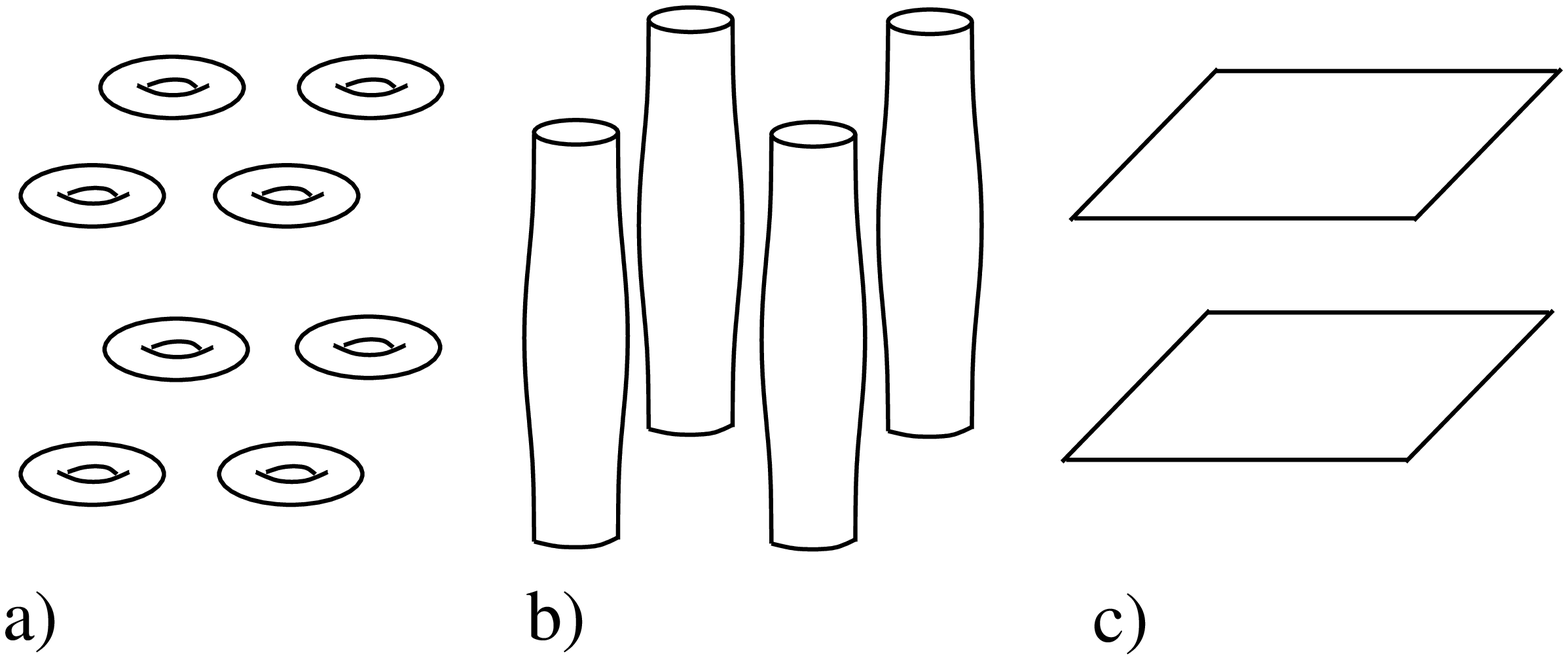,width=14.0cm,height=8cm}
\end{center}
\caption{The periodically repeated tori ${\mathbb T}^{2}$,
periodically repeated  "warped" integral
cylinders and the periodically repeated "warped"
integral planes in ${\mathbb R}^{3}$.
}
\label{tori}
\end{figure}

 Indeed, it is easy to see that all the three cases of 
periodically repeated tori ${\mathbb T}^{2}$ in
${\mathbb R}^{3}$, periodically repeated  "warped" integral
cylinders and the periodically repeated "warped"
integral planes give the topological 2-dimensional
tori ${\mathbb T}^{2}$ in 
${\mathbb T}^{3}$ after the factorization (see Fig. \ref{tori}).

 Let us note here that we call the cylinder in 
${\mathbb R}^{3}$ integral if it's axis  
is parallel to some vector of the reciprocal lattice, while the
plane in 
${\mathbb R}^{3}$ is called integral if it is generated by some
two reciprocal lattice vectors. The case $r=2$, however, has an
important difference from the cases $r=0$ and $r=1$. The matter is
that the plane in ${\mathbb R}^{3}$ is not homological to zero in 
${\mathbb T}^{3}$
(i.e. does not restrict any domain of "lower energies") after
the factorization. We can conclude so that if these planes
appear as the connected components of the physical Fermi surface
they should always come in pairs, $\Pi_{+}$ and $\Pi_{-}$,
which are parallel to each other in ${\mathbb R}^{3}$. 
The factorization  of $\Pi_{+}$ and $\Pi_{-}$ gives then the two tori
${\mathbb T}^{2}_{+}$, ${\mathbb T}^{2}_{-}$ 
with the opposite homological classes  
in ${\mathbb T}^{3}$ after the factorization. The space between the
$\Pi_{+}$ and $\Pi_{-}$ in ${\mathbb R}^{3}$ can now be taken as 
the domain
of lower (or higher) energies and the disjoint union
$\Pi_{+} \cup \Pi_{-}$ will correspond to the union
${\mathbb T}^{2}_{+} \cup {\mathbb T}^{2}_{-}$ homological to zero 
in ${\mathbb T}^{3}$.

 It can be shown that the Topological Rank of any component of
genus $2$ can not exceed $2$ also. The example of the corresponding
immersion of such component with maximal Rank is shown at 
Fig. \ref{rank}, c
and represents the two parallel planes connected by cylinders.

 At last we say that the Topological Rank of the components with
genus $g \geq 3$ can take any value $r = 0,1,2,3$.

  Let us give also the definitions of "rationality" and
"irrationality" of the direction of ${\bf B}$.

\vspace{0.5cm}

{\bf Definition 5.}

{\it Let $\{{\bf g}_{1}, {\bf g}_{2}, {\bf g}_{3}\}$ be the basis 
of the reciprocal lattice $\Gamma^{*}$. Then:
  
 1) The direction of ${\bf B}$ is rational (or has irrationality $1$)
if the numbers $({\bf B}, {\bf g}_{1})$, $({\bf B}, {\bf g}_{2})$,
$({\bf B}, {\bf g}_{3})$ are proportional to each other with
rational coefficients.
 
 2) The direction of ${\bf B}$ has irrationality $2$ if the numbers
$({\bf B}, {\bf g}_{1})$, $({\bf B}, {\bf g}_{2})$,
$({\bf B}, {\bf g}_{3})$ generate the linear space of dimension
$2$ over ${\mathbb Q}$.
    
 3) The direction of ${\bf B}$ has irrationality $3$ if the numbers
$({\bf B}, {\bf g}_{1})$, $({\bf B}, {\bf g}_{2})$,
$({\bf B}, {\bf g}_{3})$ are linearly independent over 
${\mathbb Q}$.}

\vspace{0.5cm}

 The conditions (1)-(3) can be formulated also as if the plane
$\Pi({\bf B})$ orthogonal to ${\bf B}$ contains two linearly
independent reciprocal lattice vectors, just one linearly
independent reciprocal lattice vector or no reciprocal lattice
vectors at all respectively.

 It can be seen also that if
$\{{\bf l}_{1}, {\bf l}_{2}, {\bf l}_{3}\}$ is the basis of
the original lattice in ${\bf x}$-space then the irrationality of
the direction of ${\bf B}$ will be given by the dimension of the
vector space generated by numbers

$$({\bf B}, {\bf l}_{2}, {\bf l}_{3}) \,\,\, , \,\,\,
({\bf B}, {\bf l}_{3}, {\bf l}_{1}) \,\,\, , \,\,\,
({\bf B}, {\bf l}_{1}, {\bf l}_{2}) $$
over $Q$. Easy to see that these numbers have the meanings of the
the magnetic fluxes through the faces of elementary lattice cell.

 Let us discuss now the connection between the geometry of the 
non-singular electron orbits and the topological properties of the
Fermi surface. We will briefly consider here the simple cases of
Fermi surfaces of Rank 0, 1 and 2 and come then to our basic
case of general Fermi surfaces having the maximal rank 
$r = 3$. We have then the following situations:

\vspace{0.5cm}

 1) The Fermi surface has Topological Rank $0$.

\begin{figure}
\begin{center}
\epsfig{file=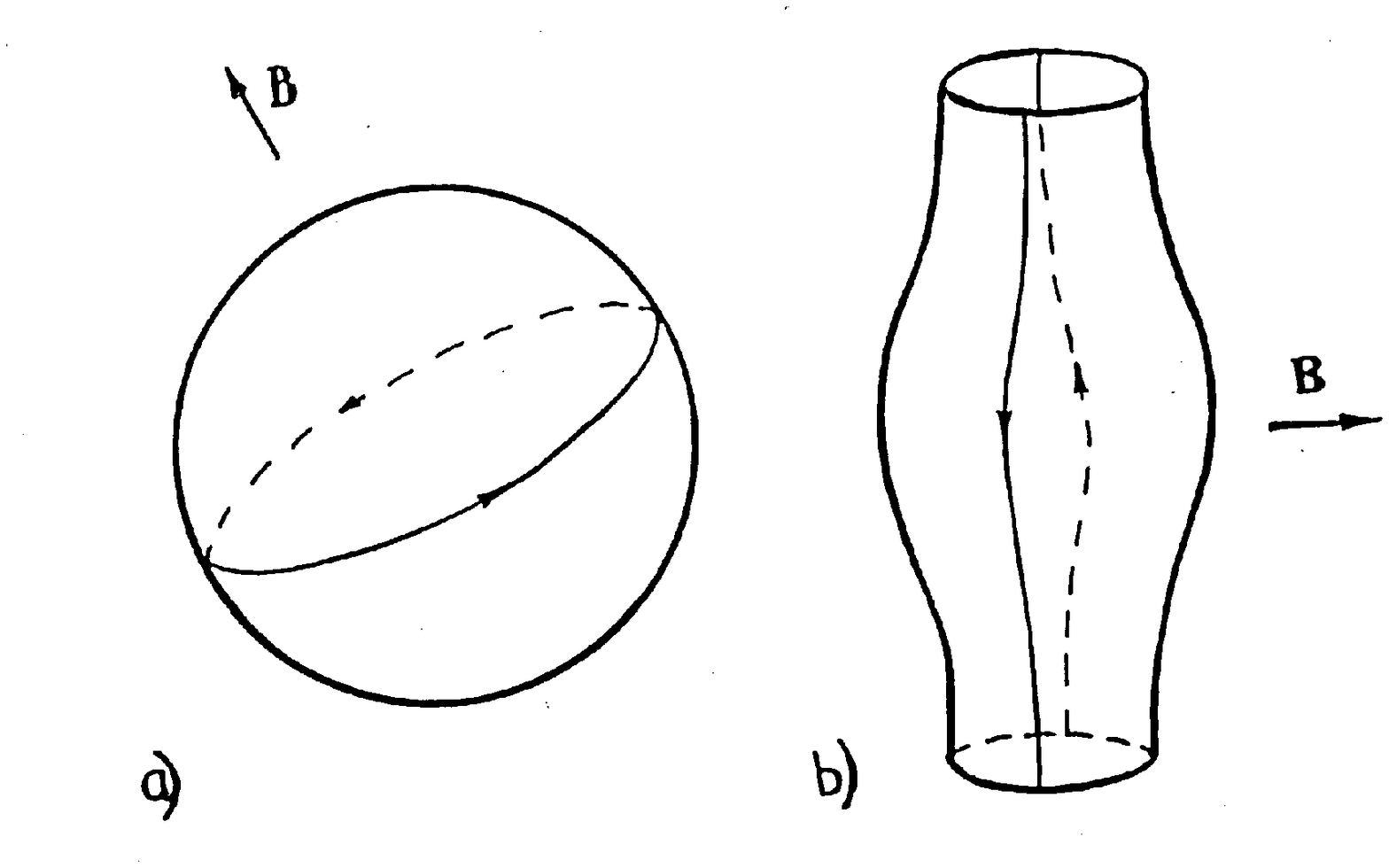,width=14.0cm,height=8cm}
\end{center}
\caption{The simplest compact and open periodic trajectories.
}
\label{clandper}
\end{figure}

Easy to see that in this simplest case all the components
of the Fermi surface are compact (Fig. \ref{clandper}, a) in 
${\mathbb R}^{3}$ and there is no open trajectories at all.

\vspace{0.5cm}

  2) The Fermi surface has Topological Rank $1$.

\begin{figure}
\begin{center}
\epsfig{file=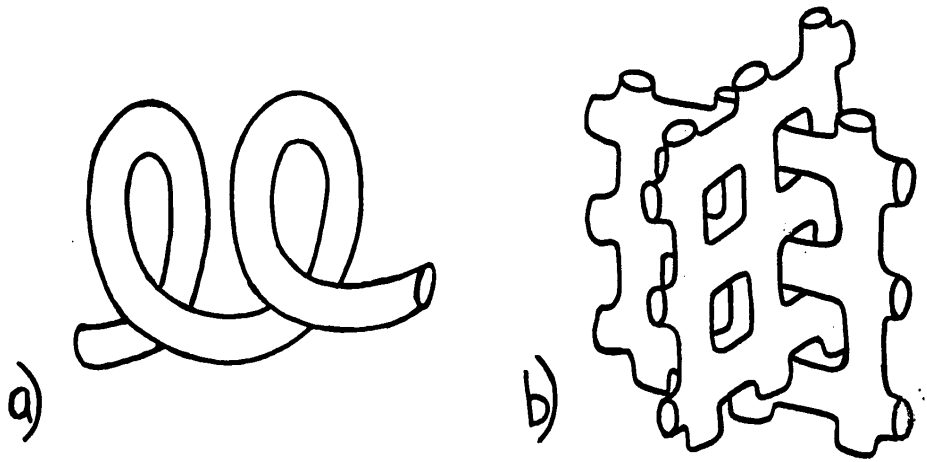,width=14.0cm,height=8cm}
\end{center}
\caption{(a) Connected component of Rank $1$ having the form of 
"helix". Open orbits are absent for any direction of ${\bf B}$.
(b) The example of the Fermi surface of Rank $2$ containing two 
components with different integral directions.
}
\label{exotic}
\end{figure}

 In this case we
can have both open and compact electron trajectories. However
the open trajectories (if they exist) should be quite simple in
this case. They can arise only if the magnetic field is orthogonal
to the mean direction of one of the components of Rank $1$ 
(periodic cylinder) and
are periodic with the same integer mean direction
(Fig. \ref{clandper}, b). There is only 
the finite number of possible mean directions of open orbits in  
this case and a finite "net" of one-dimensional curves on the  
unit sphere giving the directions of ${\bf B}$ corresponding
to the open orbits. In some special points we can have the
trajectories with different mean directions lying in different
parallel planes orthogonal to ${\bf B}$. Easy to see that in this
case the direction of ${\bf B}$ should be purely rational such
that the orthogonal plane $\Pi({\bf B})$ contains two different
reciprocal lattice vectors. It is evident also that there is only
the finite number of such directions of
${\bf B}$ clearly determined
by the mean directions of the components of Rank $1$. Let us
mention also that the existence
of open orbits is not necessary here
even for ${\bf B}$ orthogonal to the mean direction of some
component of Rank $1$ as can be seen from the example of the
"helix" represented on Fig. \ref{exotic}, a.
 Easy to see also the this type of trajectories corresponds   
to topologically integrable case according to the Definition 3 
and gives the simplest example of topologically regular
open orbit in the plane $\Pi({\bf B})$.

 Let us point out here that both cases of compact and periodic 
open trajectories were considered in details in the work 
\cite{lifazkag} where the form of the
conductivity tensor in the limit $B \rightarrow \infty$ was
obtained for these situations. Let us say also, that these
two situations are actually the only two cases where the
"dynamical" coordinates $(\epsilon, p_{z}, t)$ 
considered in \cite{lifazkag} (corresponding
to the dynamical system (\ref{dynsyst})) can be introduced
globally on the Fermi surface and the full analytic expansion
of $\sigma^{ik}(B)$ in the powers of $1/B$ can be obtained
for $\omega_{B} \tau \gg 1$. As we will see later this situation
does not take place even for more general cases of quasiperiodic
topologically regular trajectories on the Fermi surface (as well
as for more complicated "chaotic" trajectories). However, the
expressions for $\sigma^{ik}(B)$ for the case of the
open periodic trajectories give also the correct main 
approximation to $\sigma^{ik}(B)$ in the main order of
$1/B$ as we will see below. The corresponding dependence was
also assumed in the papers \cite{lifpes1,lifpes2} where the more
complicated cases of open trajectories appeared. They were also
used in \cite{novmal1,novmal2,malnov3} where the 
"Topological quantum numbers" corresponding to general 
topologically regular trajectories were introduced. As we will
see, however, the "ergodic dynamics" of (\ref{dynsyst}) in the
more general cases can be observed in the next orders in
$1/B$ in precise conductivity measurements. Let us give here
the corresponding expressions for conductivity for cases of compact 
and open periodic trajectories obtained in \cite{lifazkag}. 
We assume here that the $z$-axis is always directed along the
magnetic field ${\bf B}$ and the $x$-axis in the plane
$\Pi({\bf B})$ (orthogonal to ${\bf B}$) is directed along the
mean direction of the periodic trajectory considered in 
${\bf p}$-space. The asymptotic values of $\sigma^{ik}$
can then be written in the following for in the main order
on $1/B$ (\cite{lifazkag}):

\vspace{0.5cm} 
    
 {\bf Case 1} (compact orbits):
  
\begin{equation}
\label{sigcltr}
\sigma^{ik} \simeq {n e^{2} \tau \over m^{*}} \,
\left( \begin{array}{ccc}
(\omega_{B}\tau)^{-2} & (\omega_{B}\tau)^{-1} &
(\omega_{B}\tau)^{-1} \cr
(\omega_{B}\tau)^{-1} & (\omega_{B}\tau)^{-2} &
(\omega_{B}\tau)^{-1} \cr
(\omega_{B}\tau)^{-1} & (\omega_{B}\tau)^{-1} & *
\end{array} \right)
\end{equation}

\vspace{0.5cm}
 
 {\bf Case 2} (open periodic orbits):

\begin{equation}
\label{sigoptr}
\sigma^{ik} \simeq {n e^{2} \tau \over m^{*}} \,
\left( \begin{array}{ccc}
(\omega_{B}\tau)^{-2} & (\omega_{B}\tau)^{-1} &
(\omega_{B}\tau)^{-1} \cr
(\omega_{B}\tau)^{-1} & * & * \cr
(\omega_{B}\tau)^{-1} & * & *
\end{array} \right)
\end{equation}

 Here $\simeq$ means "of the same order in $\omega_{B}\tau$
and $*$ are some constants $\sim 1$. Let us mention also that
the relations (\ref{sigcltr})-(\ref{sigoptr}) give only the
absolute values of $\sigma^{ik}$.

\vspace{0.5cm}

 Let us mention also that the electron dynamics in ${\bf x}$-space 
can be described by additional system

$${\dot {\bf x}} = {\bf v}_{gr}({\bf p})$$
and can be easily reconstructed
for any known trajectory of (\ref{dynsyst}) in ${\bf p}$-space.
Such the $xy$-projection of any trajectory in ${\bf x}$-space
can be obtained just by rotation of corresponding trajectory
of (\ref{dynsyst}) by $\pi/2$. Easy to see then that the mean 
direction of open periodic electron trajectory in 
${\bf x}$-space coincides with the $y$-direction in our
coordinate system. The last fact can be clearly seen in the
form (\ref{sigoptr}) of the tensor $\sigma^{ik}$ as the strong
anisotropy of longitudinal conductivity in the plane $\Pi({\bf B})$
for $\omega_{B} \tau \gg 1$.

 Let us now consider the cases of Topological rank 2 and 3 of
the Fermi surface.

\vspace{0.5cm}

 3) The Fermi surface has Topological Rank 2.

 It can be easily seen that this case gives much more
possibilities for the existence of open orbits for different
directions of the magnetic field. In particular, this is the  
first case where the open orbits can exist for the generic
direction of ${\bf B}$ with irrationality $3$. So, in this case  
we can have the whole regions on the unit sphere such that the 
open orbits present for any direction of
${\bf B}$ belonging to the
corresponding region. It is easy to see, however, that the open
orbits have also a quite simple description in this case.
Namely, any open orbit (if they exist) lies in this case in   
the straight strip of the finite width for any direction of
${\bf B}$ not orthogonal to the integral planes given by the
components of Rank $2$. The boundaries of the corresponding strips
in the planes $\Pi({\bf B})$ orthogonal to ${\bf B}$ will be
given by the intersection of $\Pi({\bf B})$ with the pairs of
integral planes bounding the corresponding components of Rank $2$.
It can be also shown (\cite{dynn1}, \cite{dynn2}) that every open
orbit passes through the strip from $- \infty$ to  $+ \infty$
and can not turn back.

 According to the remarks above the contribution to the conductivity
given by every family of orbits with the same mean direction 
reveals the strong anisotropy when $\omega_{B}\tau \rightarrow \infty$
and coincides in the main order with the formula (\ref{sigoptr})
for the open periodic trajectories.

 For purely rational directions of ${\bf B}$ we can have the  
situation when the open trajectories with different mean  
directions present on different components of the Fermi surface. 
For example, for the "exotic" surface shown at 
Fig. \ref{exotic}, b 
we will have the periodic trajectories along both the $x$ and $y$
directions in different planes orthogonal to ${\bf B}$ if
${\bf B}$ is directed along the $z$ axis. However, it can be   
shown that for any direction of ${\bf B}$ which is not purely
rational this situation is impossible. We have so that for any
direction of ${\bf B}$ with irrationality $2$ or $3$ all the
open orbits will have the same mean direction and can exist 
only on the components of Rank $2$ with the same (parallel)
integral orientation. This statement is a corollary
of more general topological theorem which we will discuss
below.

 At last we note that the directions of ${\bf B}$ orthogonal 
to one of the components of Rank $2$ are purely rational and
all the non-singular open orbits (if they exist) are rational
periodic in this case.
For any family of such orbits with the same mean direction   
the corresponding contribution to the
conductivity can then be written in the form (\ref{sigoptr})
in the appropriate coordinate system. However, the direction
of open orbits can not be predicted apriori in this case. 

  Let us say that the trajectories of this type have already
all the features of the general topologically integrable
situation and the topologically regular trajectories appearing
in the more complicated case of topological rank 3 have
in fact very similar structure to the described above. 
However, we will need the special procedure of topological
reconstruction preserving all the open trajectories on the
Fermi surface which permits to see this fact in more complicated
general situation. We will consider here in details also the
"ergodic" properties of such trajectories on the corresponding 
pieces of the Fermi surface using the topological structure 
of this kind.

 Let us start now with the most general and complicated case
of arbitrary Fermi surface of Topological rank 3.

 We describe first the convenient procedure
(\cite{dynn4},\cite{dynn7}) of reconstruction of the constant
energy surface when the direction of ${\bf B}$ is fixed.

\begin{figure}
\begin{center}
\epsfig{file=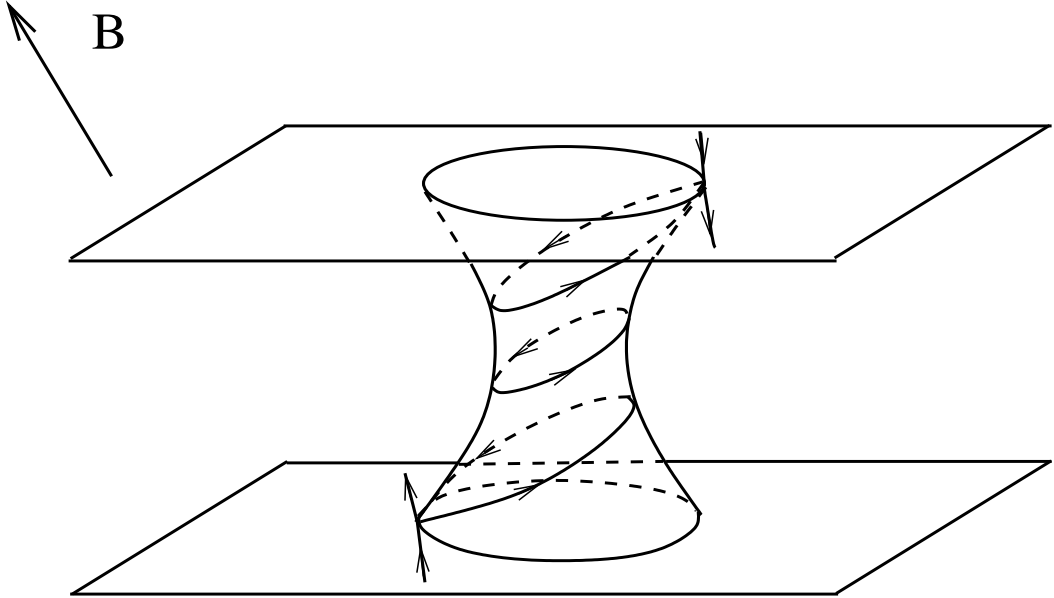,width=14.0cm,height=8cm}
\end{center}
\caption{The cylinder of compact trajectories bounded by the singular
orbits.(The simplest case of just one critical point on the singular
trajectory.)
}
\label{cylcltr}
\end{figure}

 As we said already, we admit only the non-degenerate
singularities of the dynamical system (\ref{dynsyst}) having
the form of the non-degenerate poles or non-degenerate
saddle points. The singular trajectories passing through the
critical points (and the critical points themselves) divide
the set of trajectories into the different parts corresponding
to different types of trajectories on the Fermi surface. As
we also already said, we will not be interested here in the
geometry of compact electron orbit in the "geometric limit"
$\omega_{B} \tau \rightarrow \infty$. It's not difficult
to show that the pieces of the Fermi surface carrying the
compact orbits can be either infinite or finite cylinders in 
${\mathbb R}^{3}$ bounded by the
singular trajectories (some of them maybe just points of minimum
or maximum) at the bottom and at the top (see Fig. \ref{cylcltr}).

 Easy to see that the first case corresponds then to the whole 
component of rank 1 carrying just the compact trajectories while the 
second case gives the part of more complicated Fermi surface
filled by the compact trajectories.

 Let us remove now all the parts containing the non-singular
compact trajectories from the Fermi surface. The 
remaining part

$$S_{F}/({\rm Compact \, Nonsingular \, Trajectories}) \,\,\, =
\,\,\, \cup_{j} \, S_{j}$$
is a union of the $2$-manifolds $S_{j}$ with boundaries
$\partial S_{j}$ who are the compact singular trajectories.
The generic type in this case is a separatrix orbit with just one
critical point like on the Fig. \ref{cylcltr}.

 Easy to see that the open orbit will not be affected at all
by the construction described above and the rest of the Fermi
surface gives the same open orbits as all possible intersections
with different planes orthogonal to ${\bf B}$.

\vspace{0.5cm}

{\bf Definition 6.}

{\it We call every piece $S_{j}$ the
{\bf "Carrier of open trajectories"}. The trajectory is "chaotic"
if the genus $g(S_{j})$ is greater than $1$. The case
$g(S_{j}) = 1$ we call "Topologically Completely Integrable".}
\footnote{Such systems on ${\mathbb T}^{2}$ were discussed for 
example in \cite{arnold}; the generic open orbits
are topologically equivalent to the straight
lines. Ergodic properties of such systems indeed can be nontrivial
as it was found by Ya.Sinai and K.Khanin in \cite{sinkhan}.}
 
\vspace{0.5cm}

\begin{figure}
\begin{center}
\epsfig{file=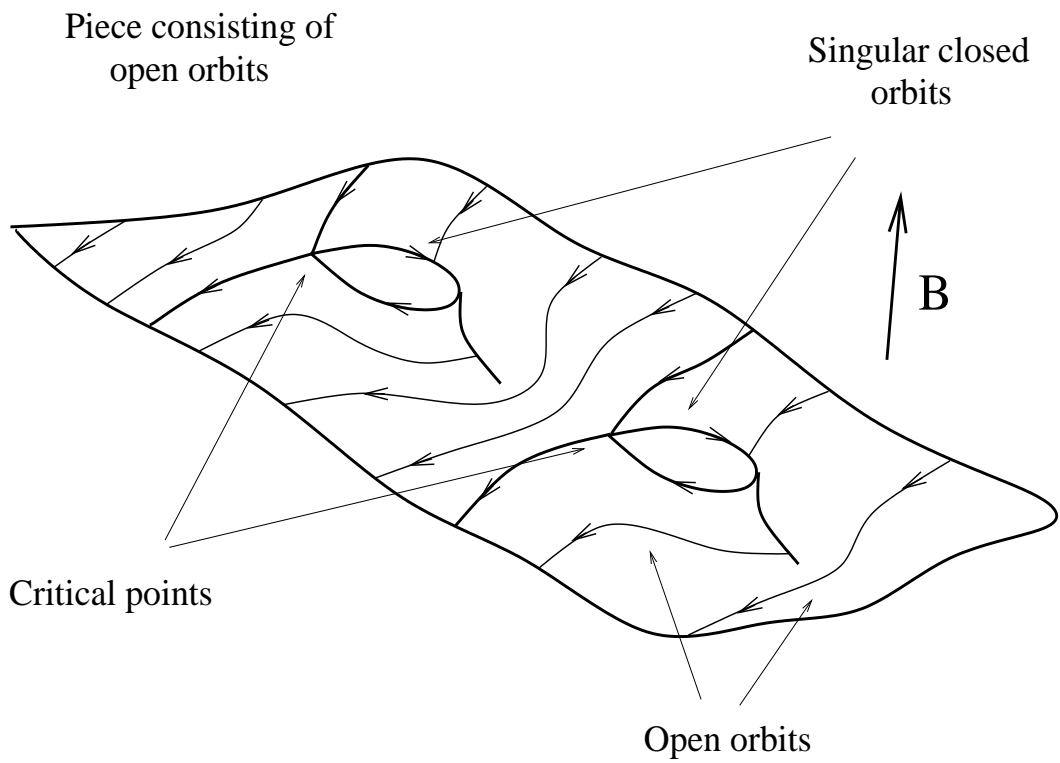,width=14.0cm,height=10cm}
\end{center}
\caption{The reconstructed constant energy surface
with removed compact orbits the two-dimensional discs attached
to the singular orbits in the generic case of just one critical
point on every singular orbit.
}
\label{reconst}
\end{figure}

 Let us fill in the holes by
topological $2D$ discs lying in the planes orthogonal to ${\bf B}$
and get the closed surfaces

$${\bar S}_{j} \,\,\, = \,\,\, S_{j} \cup (2-discs)$$
(see Fig. \ref{reconst}).

 This procedure gives again the periodic surface
${\bar S}_{\epsilon}$ after the reconstruction and we can define  
the "compactified carriers of open trajectories" both in 
${\mathbb R}^{3}$ and ${\mathbb T}^{3}$.

 Let us formulate now the main topological theorems concerning
the geometry of open trajectories which made a breakthrough
in the theory of such dynamical systems on the Fermi surfaces
(\cite{zorich}, \cite{dynn3}).

\vspace{0.5cm}

{\bf Theorem $1$.} \cite{zorich}

{\it Let us fix the energy level $S_{\epsilon}$ and any rational
direction ${\bf B}_{0}$ such that no two saddle points on
$S_{\epsilon}$ are connected in 
${\mathbb R}^{3}$ by the singular electron
trajectory. Then for all the directions of ${\bf B}$
close enough to ${\bf B}_{0}$ every open
trajectory lies in the strip of the finite width between two
parallel lines in the plane orthogonal to ${\bf B}$.
}

\vspace{0.5cm}

 In fact, the proof of the Theorem $1$ was based on
the statement that genus of every compactified carrier of open
orbits ${\bar S}_{j}$ is equal to $1$ in this case.
 
\vspace{0.5cm}

{\bf Theorem $2$.} \cite{dynn3}

{\it Let a generic dispersion relation

$$\epsilon({\bf p}): \,\,\, 
{\mathbb T}^{3} \,\,\, \rightarrow {\mathbb R}$$   
be given such that for level
$\epsilon({\bf p}) = \epsilon_{0}$ the genus $g$ of some
carrier of open trajectories ${\bar S}_{i}$ is greater than $1$.
Then there exists an open interval
$(\epsilon_{1}, \epsilon_{2})$ containing $\epsilon_{0}$
such that for all $\epsilon \neq \epsilon_{0}$ in this interval 
the genus of carrier of open trajectories is less than $g$.
}

\vspace{0.5cm}

 Let us note now that the genus 1 corresponds precisely
to the two-dimensional tori ${\mathbb T}^{2}$ embedded to the
three dimensional torus ${\mathbb T}^{3}$ as the compactified
carriers of open trajectories on the Fermi surface. The
corresponding components in the covering space can thus be
either the periodically deformed cylinders or the periodically
deformed integral plane in ${\mathbb R}^{3}$ filled by the 
open orbits. Easy to see then that the first case can arise
only for the directions of ${\bf B}$ orthogonal to some
integer vector in ${\mathbb R}^{3}$ (i.e. to the mean direction
of the corresponding component carrying open orbits) and
corresponds to the periodic open orbits in $\Pi({\bf B})$.
The second case corresponds to the more general situation
and can arise for generic directions of ${\bf B}$ being
stable with respect to all small rotations of the magnetic
field. Easy to see the direct analogy of this situation with
the cases of Fermi surfaces of rank 1 and 2 respectively.
Actually, we show in this construction that for the fixed direction
of ${\bf B}$ we can replace the Fermi surface by some other
surface having the simpler structure which gives us the same open
trajectories in all planes $\Pi({\bf B})$ orthogonal to the 
magnetic field. As can be easily seen also, all the trajectories
can correspond only to topologically integrable case if the
carriers of open trajectories all have the genus 1. The mean
direction of generic open trajectories can be obtained then
as the intersections of the planes orthogonal to ${\bf B}$
with the integral planes in reciprocal lattice represented
by the periodically deformed integral planes carrying open orbits.

 The Theorem $2$ claims then that only the
"Topologically Integrable case" can be stable with
respect to the small variations of energy level and has a
generic properties in this situation. It follows also from  
Theorem 1 that this case is the only stable case with respect to the
small rotations of ${\bf B}$ and then should be considered as the
generic situation among all the situations when the non-compact
electron trajectories arise on the Fermi surface. 

 The important property of the compactified components of genus 1
arising for the generic directions of ${\bf B}$ is following:
they are all  parallel in average in ${\mathbb R}^{3}$ 
and do not intersect each other. 
This property, playing the crucial role for conductivity
phenomena, was first pointed out in \cite{novmal1} and called later
the "{\bf Topological resonance}" for the system (\ref{dynsyst}).
According to this property, all the stable topologically regular
open orbits in all planes orthogonal to ${\bf B}$ have the same
mean direction and then give the same form (\ref{sigoptr}) of
contribution to conductivity in the appropriate coordinate system
common for all of them. This fact gives the experimental possibility
to measure the mean direction of non-compact topologically regular
orbits both in ${\bf x}$ and ${\bf p}$ spaces from the anisotropy
of conductivity tensor $\sigma^{ik}$. From the physical point of
view, all the regions on the unit sphere 
(giving the directions of magnetic field) where the stable open
orbits exist can be represented as the "stability zones"
$\Omega_{\alpha}$ such that 
each zone corresponds to some integral plane $\Gamma_{\alpha}$ 
common to all the points of stability zone $\Omega_{\alpha}$.
The plane $\Gamma_{\alpha}$ is then the integral plane in 
reciprocal lattice given by the
integral mean direction of the components of genus 1 carrying
open orbits and defining the mean directions of open orbits
in ${\bf p}$-space
for any direction of ${\bf B}$ belonging to $\Omega_{\alpha}$
just as the intersection with the plane orthogonal to ${\bf B}$.
As can be easily seen from the form of (\ref{sigoptr}) this
direction always coincides with the unique direction in 
${\mathbb R}^{3}$ corresponding to the decreasing of 
longitudinal conductivity as $\omega_{B} \tau \rightarrow \infty$.

  The corresponding integral planes $\Gamma_{\alpha}$ can then
be given by three integer numbers
$(n^{1}_{\alpha}, n^{2}_{\alpha}, n^{3}_{\alpha})$
(up to the common multiplier) from the equation  
$$n^{1}_{\alpha} [{\bf x}]_{1} + n^{2}_{\alpha} [{\bf x}]_{2} +
n^{3}_{\alpha} [{\bf x}]_{3} = 0$$
where $[{\bf x}]_{i}$ are the coordinates in the basis
$\{{\bf g}_{1}, {\bf g}_{2}, {\bf g}_{3}\}$ of the reciprocal
lattice, or equivalently

$$n^{1}_{\alpha} ({\bf x}, {\bf l}_{1}) +
n^{2}_{\alpha} ({\bf x}, {\bf l}_{2}) +
n^{3}_{\alpha} ({\bf x}, {\bf l}_{3}) = 0$$
where $\{{\bf l}_{1}, {\bf l}_{2}, {\bf l}_{3}\}$ is the basis
of the initial lattice in the coordinate space.

 We see then that the direction of conductivity decreasing 
${\hat \eta} = (\eta_{1}, \eta_{2}, \eta_{3})$ satisfies to relation

$$n^{1}_{\alpha} ({\hat \eta}, {\bf l}_{1}) +
n^{2}_{\alpha} ({\hat \eta}, {\bf l}_{2}) +
n^{3}_{\alpha} ({\hat \eta}, {\bf l}_{3}) = 0$$
for all the points of stability zone $\Omega_{\alpha}$ which
makes possible the experimental observation of numbers 
$(n^{1}_{\alpha}, n^{2}_{\alpha}, n^{3}_{\alpha})$.

 The numbers $(n^{1}_{\alpha}, n^{2}_{\alpha}, n^{3}_{\alpha})$
were called in \cite{novmal1} the "Topological Quantum numbers" 
of a dispersion relation in metal.

 Let us add also that the number of tori 
${\mathbb T}_{i}^{2}$ being even
can still be different for the different points of stability
zone $\Omega_{\alpha}$. We can then introduce in the general
situation the "sub-boundaries" of the stability zone which are
the piecewise smooth curves inside $\Omega_{\alpha}$ where   
the number of tori generically changes by $2$.
The asymptotic behavior of conductivity will still be
described by the formula (\ref{sigoptr}) in this case
but the dimensionless coefficients will then "jump"
on the sub-boundaries of stability zone. Let us however
mention here that this situation can be observed only for
rather complicated Fermi surfaces.

\vspace{0.5cm}

 Let us say now some words about the special situations when
the carriers of open trajectories have genus more than 1
which can also happen for rather complicated Fermi surfaces.

 It was first shown by S.P.Tsarev (\cite{tsarev}) the
more complicated chaotic open orbits can still exist on
rather complicated Fermi surfaces $S_{F}$. Such, the   
example of open trajectory which does not lie in any finite
strip of finite width was constructed. The corresponding
direction of ${\bf B}$ had the irrationality $2$ in this 
example and the closure of the open orbit was a "half" of    
the surface of genus $3$ separated by the singular
closed trajectory non-homotopic to zero in 
${\mathbb T}^{3}$. However,
the trajectory had in this case the asymptotic direction
even not being restricted by any straight strip of finite
width in the plane orthogonal to ${\bf B}$.

 As was shown later in \cite{dynn4} this situation
always takes place for any chaotic trajectory for the
directions of ${\bf B}$ with irrationality $2$. We have so,
that for non-generic "partly rational" directions of ${\bf B}$
the chaotic behavior is still not "very complicated" 
and resembles some features of stable open electron trajectories.

  The corresponding asymptotic behavior of conductivity should
reveal also the strong anisotropy properties in the plane
orthogonal to ${\bf B}$ although the exact form of
$\sigma^{ik}$ will be slightly different from (\ref{sigoptr}) for
this type of trajectories. By the same reason, the asymptotic
direction of orbit can be measured experimentally in this case
as the direction of lowest longitudinal conductivity in
${\mathbb R}^{3}$ according to kinetic theory. The measure of the
corresponding set on the unit sphere
is obviously zero for such type of trajectories
being restricted by the measure of directions of irrationality
$2$. We will consider here the Tsarev example in more details
below and discuss the corresponding conductivity behavior 
for $B \rightarrow \infty$.

 The more complicated examples of chaotic open orbits were
constructed in \cite{dynn4} for
the Fermi surface having genus $3$.
The direction of the magnetic field has the irrationality $3$
in this case and the closure of the chaotic trajectory
covers the whole Fermi surface in 
${\mathbb T}^{3}$. These types of the
open orbits do not have any asymptotic direction in the planes
orthogonal to ${\bf B}$ and have rather complicated form
"walking everywhere" in these planes. Let us also discuss later
this case in more details.

\vspace{0.5cm}
 
 Let us formulate now the recent topological results concerning
the general behavior of quasiclassical electron orbits on the
energy levels of the generic dispersion relation 
$\epsilon = \epsilon({\bf p})$ on ${\mathbb T}^{3}$.

 The systematic investigation of the open orbits was completed
after the works \cite{zorich,dynn3,novmal1} by I.A.Dynnikov 
(see \cite{dynn4,dynn7}). In particular the total picture of 
different types of the open orbits for generic dispersion relations 
was presented in \cite{dynn7}. Let us just formulate here the
main results of \cite{dynn4,dynn7} in the form of Theorem.

\vspace{0.5cm}

{\bf Theorem $3$} (\cite{dynn4}, \cite{dynn7}).

{\it Let us fix the dispersion relation
$\epsilon = \epsilon({\bf p})$
and the direction of ${\bf B}$ of irrationality $3$ and consider
all the energy levels for
$\epsilon_{min} \leq \epsilon \leq \epsilon_{max}$. Then:

 1) The open electron trajectories exist for all the energy values
$\epsilon$ belonging to the closed connected energy interval
$\epsilon_{1}({\bf B}) \leq \epsilon \leq \epsilon_{2}({\bf B})$
which can degenerate to just one energy level
$\epsilon_{1}({\bf B}) = \epsilon_{2}({\bf B}) =
\epsilon_{0}({\bf B})$.

 2) For the case of the nontrivial energy interval the set of
compactified carriers of open trajectories  
${\bar S}_{\epsilon}$ is always a disjoint union of
two-dimensional tori ${\mathbb T}^{2}$ in 
${\mathbb T}^{3}$ for all
$\epsilon_{1}({\bf B}) \leq \epsilon \leq \epsilon_{2}({\bf B})$.
All the tori 
${\mathbb T}^{2}$ for all the energy levels do not intersect
each other and have the same (up to the sign) indivisible
homology class 
$c \in H_{2}({\mathbb T}^{3},{\mathbb Z})$, $c \neq 0$.
The number of tori 
${\mathbb T}^{2}$ is even for every fixed energy level
and the corresponding covering ${\bar S}_{\epsilon}$ in 
${\mathbb R}^{3}$
is a locally stable family of parallel ("warped") integral planes
$\Pi^{2}_{i} \subset {\mathbb R}^{3}$ 
with common direction given by $c$.
The form of ${\bar S}_{\epsilon}$ described above is locally
stable with the same homology class 
$c \in H_{2}({\mathbb T}^{3})$ under  
small rotations of ${\bf B}$.
All the open electron trajectories at all the energy levels
lie in the strips of finite width with the same direction and
pass through them. The mean direction of the trajectories is given
by the intersections of planes $\Pi({\bf B})$ with the integral
family $\Pi^{2}_{i}$ for the corresponding "stability zone" on  
the unit sphere.

 3) The functions $\epsilon_{1}({\bf B})$, $\epsilon_{2}({\bf B})$
defined for the directions of ${\bf B}$ of irrationality $3$ can
be continuated on the unit sphere $S^{2}$ as the piecewise smooth
functions such that
$\epsilon_{1}({\bf B}) \geq \epsilon_{2}({\bf B})$ everywhere
on the unit sphere.

 4) For the case of trivial energy interval
$\epsilon_{1} = \epsilon_{2} = \epsilon_{0}$ the corresponding
open trajectories may be chaotic. Carrier of the chaotic open
trajectory is homologous to zero in 
$H_{2}({\mathbb T}^{3},{\mathbb Z})$ and has genus
$\geq 3$. For the generic energy level $\epsilon = \epsilon_{0}$
the corresponding directions of magnetic fields 
belong to the countable union of the codimension $1$ subsets. 
Therefore a measure of this set is equal to zero on $S^{2}$.
}

\vspace{0.5cm}

 The whole manifold ${\bar S}_{\epsilon}$ is always homologous to 
zero in ${\mathbb T}^{3}$ and all the two dimensional tori 
${\mathbb T}^{2}$ can be
always divided in two equal groups 
$\{{\mathbb T}^{2}_{i+}\}$, $\{{\mathbb T}^{2}_{i-}\}$ 
according to the direction of the electron motion.
As can be proved using Theorem $1$ (\cite{dynn7}) the "stability
zones" form the everywhere dense set on the unit sphere for the   
generic dispersion relations. All the non-compact trajectories   
stable under the small rotations of ${\bf B}$ should have thus   
the form described above.

 Let us however remind here that for the conductivity phenomena
in metals only the Fermi level $\epsilon({\bf p}) = \epsilon_{F}$
is important and we can not get any information about stability
zones of dispersion relation such that 
$\epsilon_{F} \notin [\epsilon_{2}({\bf B}), \epsilon_{1}({\bf B})]$.

 Let us say also that Theorem 3 describes the generic situation 
of the directions of ${\bf B}$ of irrationality 3 and some
additional features can arise for the purely (irrationality 1)
or "partly" (irrationality 2) rational directions of ${\bf B}$.
We will not discuss here all these features in details and just
make a reference on the articles \cite{novmal2,malnov3} where
the general situation was considered.

\section{The behavior of conductivity in the case of topologically 
regular open orbits.}

 Let us now discuss briefly the conductivity behavior in
"geometric strong magnetic field limit". We prove below the analog
of Lifshitz-Azbel-Kaganov-Peschanskii formulae in this case
(\cite{lifazkag,lifpes1,lifpes2}).
We will start with generic
"topologically regular" open orbits having the "strong asymptotic
directions in ${\mathbb R}^{3}$.
 
 Let us come back to the formula (\ref{sigmaik}) and consider
the case of the open orbits having the form shown at 
Fig. \ref{regandch}, a. Let us take the $x$-axis in 
${\mathbb R}^{3}$ with the direction corresponding to the mean 
directions of open orbits in ${\bf p}$-space and the $z$-axis
directed along the magnetic field ${\bf B}$. The projection
of trajectory on the plane orthogonal to ${\bf B}$ will have 
the same form in ${\bf x}$-space just rotated by $\pi/2$ such that
the mean direction of open orbits will be directed along
the $y$-axis in our coordinate system. 

 We can write then that generally

$$\langle v^{x}_{gr} \rangle = 0 \,\,\, , \,\,\,
\langle v^{y}_{gr} \rangle \neq 0 \,\,\, , \,\,\,
\langle v^{z}_{gr} \rangle \neq 0 $$
on these trajectories, where $\langle \dots \rangle$ means the
averaging along the trajectory according to system 
(\ref{dynsyst}).

 Moreover we can claim that the function

$$p^{y}({\bf p},-t) - p^{y}({\bf p},0) =  
{eB \over c} 
\int_{0}^{t} v^{x}_{gr} ({\bf p}, - t^{\prime}) d t^{\prime} $$
is a bounded function for any trajectory shown at 
Fig. \ref{regandch}, a, as follows from its geometric form.

 Let us consider now the components $\sigma^{i1}$ 
(or $\sigma^{1i}$) of the conductivity tensor 
$\sigma^{ik}({\bf B})$. Using the integration by parts
we can write:

$$\sigma^{i1} ({\bf B}) \, = \, 
e^{2} {c \over eB\tau} \int\dots\int 
\left( - {\partial f_{T} \over \partial \epsilon} \right) 
v^{i}_{gr}({\bf p}) \times \hspace{5cm}$$
\begin{equation}
\label{sigform}
\hspace{5cm} \times \left[ \int_{0}^{+\infty}
\left(p^{y}({\bf p},-t) - p^{y}({\bf p},0)\right) e^{-t/\tau} 
dt \right] {d^{3}p \over (2\pi\hbar)^{3}}
\end{equation}

 For the component $\sigma^{11}$ we can use then the formula

$$\sigma^{11} ({\bf B}) \, = \,
e^{2} {c^{2} \over (eB)^{2}\tau}
\int\dots\int 
\left( - {\partial f_{T} \over \partial \epsilon} \right)
\times \hspace{4cm}$$
$$\hspace{4cm} \times \left[ \int_{0}^{+\infty} {1 \over 2} 
{d (p^{y}({\bf p},-t) - p^{y}({\bf p},0))^{2} \over dt}
e^{-t/\tau} dt \right] {d^{3}p \over (2\pi\hbar)^{3}} = $$

$$= e^{2} \left( {c \over eB\tau} \right)^{2} \int\dots\int
\left( - {\partial f_{T} \over \partial \epsilon} \right)
\left[ \int_{0}^{+\infty} {1 \over 2}
\left(p^{y}({\bf p},-t) - p^{y}({\bf p},0)\right)^{2}
e^{-t/\tau} dt \right] {d^{3}p \over (2\pi\hbar)^{3}}$$

 The integral 

$$I = \int_{0}^{+\infty} {1 \over 2}
\left(p^{y}({\bf p},-t) - p^{y}({\bf p},0)\right)^{2}
e^{-t/\tau} dt$$
can be evaluated as $I \leq 1/2 \tau P_{0}^{2}$ where
$P_{0}$ is the common constant bounding the values 
$|p^{y}({\bf p},-t) - p^{y}({\bf p},0)|$ for all the regular 
trajectories (the width of straight strip). The value
$\sigma^{11} ({\bf B})$ can then be written as

\begin{equation}
\label{sigma11}
\sigma^{11} ({\bf B}) \sim {n e^{2} \tau \over m^{*}}
{\alpha \over (\omega_{B}\tau)^{2}} + o \left(
{1 \over (\omega_{B}\tau)^{2}}\right) 
\end{equation}
when $B \rightarrow \infty$.

 The value of constant $\alpha$ can then be defined in 
terms of the integral
$${1 \over 2} \int\int \langle
\left(p^{y}({\bf p},-t) - p^{y}({\bf p},0)\right)^{2} \rangle
{d^{2} S \over (2\pi\hbar)^{3} v_{gr}({\bf p})} $$
over the Fermi surface and is proportional to the square of
effective width of straight lines bounding the topologically
regular trajectories.

 Let us say now that the total expansion in $(1/B)$ can not
be actually made for the generic regular trajectories because
of their ergodic behavior. However, the difference between
the quasiperiodic and periodic trajectories does not appear in 
the main terms of the formula (\ref{sigoptr}) and can be detected 
only in the next smaller approximations for $B \rightarrow \infty$.
This fact permits actually to use the formula (\ref{sigoptr})
also for regular quasiperiodic trajectories when speaking just
about the geometric properties of $\sigma^{ik}$. Let us add
also that the next approximations to (\ref{sigoptr}) depend
on whether the saddle critical points really present on the
carriers of open orbits or not.

 For the components $\sigma^{i1}$ or $\sigma^{1i}$, $i = 2,3$
we can not use the second integration by parts and we should 
put then

\begin{equation}
\label{sigmai1}
\sigma^{i1}({\bf B}) \sim \sigma^{1i}({\bf B}) \sim
{n e^{2} \tau \over m^{*}}
{\beta^{i} \over \omega_{B}\tau} +
o \left( {1 \over \omega_{B}\tau}\right) 
\end{equation}
according to the formula (\ref{sigform}). 

 Again the same remarks about next approximations in 
$\sigma^{i1}$ and $\sigma^{1i}$ can also be made in this case.

 Let us consider now components $\sigma^{ik}$ where both
$i,j \neq 1$. We can write the the formula (\ref{sigmaik})
using the averaged values of $v^{y}_{gr}$ and $v^{z}_{gr}$
in the following way:

$$\sigma^{ik }({\bf B}) = e^{2} \tau \int\dots\int
\left( - {\partial f_{T} \over \partial \epsilon} \right)
\langle v^{i}_{gr} \rangle \langle v^{k}_{gr} \rangle
{d^{3}p \over (2\pi\hbar)^{3}} + $$

$$+ e^{2} \int\dots\int
\left( - {\partial f_{T} \over \partial \epsilon} \right)
\left( v^{i}_{gr}({\bf p}) - \langle v^{i}_{gr} \rangle \right)
\times \hspace{5cm} $$

\begin{equation}
\label{condavf}
\hspace{5cm} \times \left[ \int_{0}^{+\infty} 
\left( v^{k}_{gr}({\bf p}, - t) - 
\langle v^{k}_{gr} \rangle \right)
e^{-t/\tau} dt \right] {d^{3}p \over (2\pi\hbar)^{3}}
\end{equation}
(It's not difficult to see the the "cross terms" will disappear 
after the integration over ${\mathbb T}^{3}$.)

 The first part of this formula gives the finite values for
$\sigma^{ik}({\bf B})$ when $B \rightarrow \infty$ while the 
second becomes 
zero in the same limit. Easy to see that the last statement
is just a simple corollary of the fact that the integral
with decreasing exponent approaches the mean value on the 
trajectory in the limit $B \rightarrow \infty$ 
(up to the factor $1/\tau$) according to the system (\ref{dynsyst}).
Let us say again, however, that  
the $B$-dependence is more complicated in general here 
than in the case of purely periodic trajectories
and can not be generically expanded in the powers of $1/B$.

 Thus we can see that the formula (\ref{sigoptr}) gives the main
part of $\sigma^{ik}({\bf B})$ also for general topologically regular 
trajectories with irrational mean directions. The cases of compact
and purely periodic trajectories, however, are characterized by the
existence of full analytic expansions in $1/B$ which do exist in
general for more complicated quasiperiodic orbits.

 Let us make also some additional remark on the periodic and 
quasiperiodic open trajectories. Namely, the periodic open orbits
arise actually in every "stability zone" when the intersection 
of the carrier of open orbits with the plane $\Pi({\bf B})$ has
the rational mean direction. Easy to see that this situation
can appear if the direction of ${\bf B}$ has irrationality 2
and always arise for the purely rational directions of ${\bf B}$.
This actually means that the corresponding orbits in 
${\mathbb T}^{3}$ are not everywhere dense on the tori 
${\mathbb T}^{2}_{i}$ (defined for a given stability zone)
but become closed (in ${\mathbb T}^{3}$) unlike the situation
with irrational mean direction of open orbits. The mean values of
$v^{i}_{gr}$ then actually depend on the open periodic trajectory
and do not coincide with the value of $v^{i}_{gr}$ averaged
over the carrier of open trajectory. We know just that averaging
of these mean values over all the periodic trajectories will
give us the values of $\langle v^{i}_{gr} \rangle$ close to the 
same values for close irrational directions of ${\bf B}$.
However, the limit $B \rightarrow \infty$ for tensor 
$\sigma^{ik}({\bf B})$ ($i,k = 2,3$) will be given by formula

$$\sigma^{ik} = e^{2} \tau \int\dots\int
\left( - {\partial f_{T} \over \partial \epsilon} \right)
\langle v^{i}_{gr} \rangle ({\bf p}) 
\langle v^{k}_{gr} \rangle ({\bf p})
{d^{3}p \over (2\pi\hbar)^{3}}$$
containing the integration of products 
$\langle v^{i}_{gr} \rangle ({\bf p})
\langle v^{k}_{gr} \rangle ({\bf p})$
which can not be replaced by the "global" mean values
$\langle v^{i}_{gr} \rangle$, $\langle v^{k}_{gr} \rangle$.
As a corollary the corresponding values of $\sigma^{ik}$
will be actually different in this limit from the case of 
purely irrational directions of ${\bf B}$. Such we can claim that 
the longitudinal conductivities $\sigma^{22}$ and $\sigma^{33}$
should be actually bigger for $B \rightarrow \infty$
than the same values for close irrational directions of ${\bf B}$.

 We can see then that except the smooth dependence on the direction 
of ${\bf B}$ (connected with the dependence on ${\bf B}$ of total
phase volume of open trajectories) we will have also the "sharp" 
peaks in the conductivity for the rational mean directions of open
orbits. The corresponding "peaks" however will be quite small for
the rational directions with big denominators. Easy to see also that
this behavior does not affect the "Topological characteristics"
of conductivity introduced above.

\section{The chaotic cases.}

\begin{figure}
\begin{center}
\epsfig{file=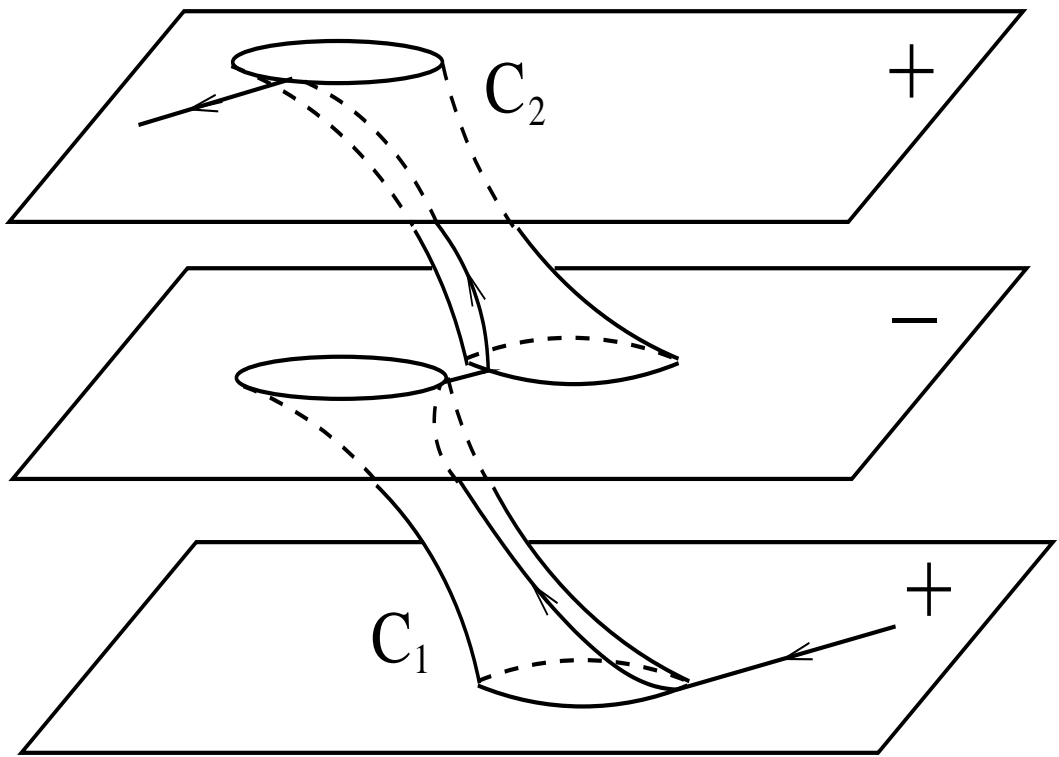,width=14.0cm,height=10cm}
\end{center}
\caption{The Fermi surface for Tsarev's example of chaotic 
trajectory.
}
\label{tsarex1}
\end{figure}

 Let us say now some words about chaotic trajectories which can 
arise in the special cases for rather complicated Fermi surfaces.
We will first describe the Tsarev's example of chaotic trajectory 
having an asymptotic direction in ${\mathbb R}^{3}$ 
(\cite{tsarev}). Let us consider the Fermi surface $S_{F}$
consisting of two sets of horizontal parallel layers 
$\{Q_{+}\}$ and $\{Q_{-}\}$ connected by 
two sets of inclined cylinders $\{C_{1}\}$, $\{C_{2}\}$
in ${\mathbb R}^{3}$ (see Fig. \ref{tsarex1}).

\begin{figure}
\begin{center}
\epsfig{file=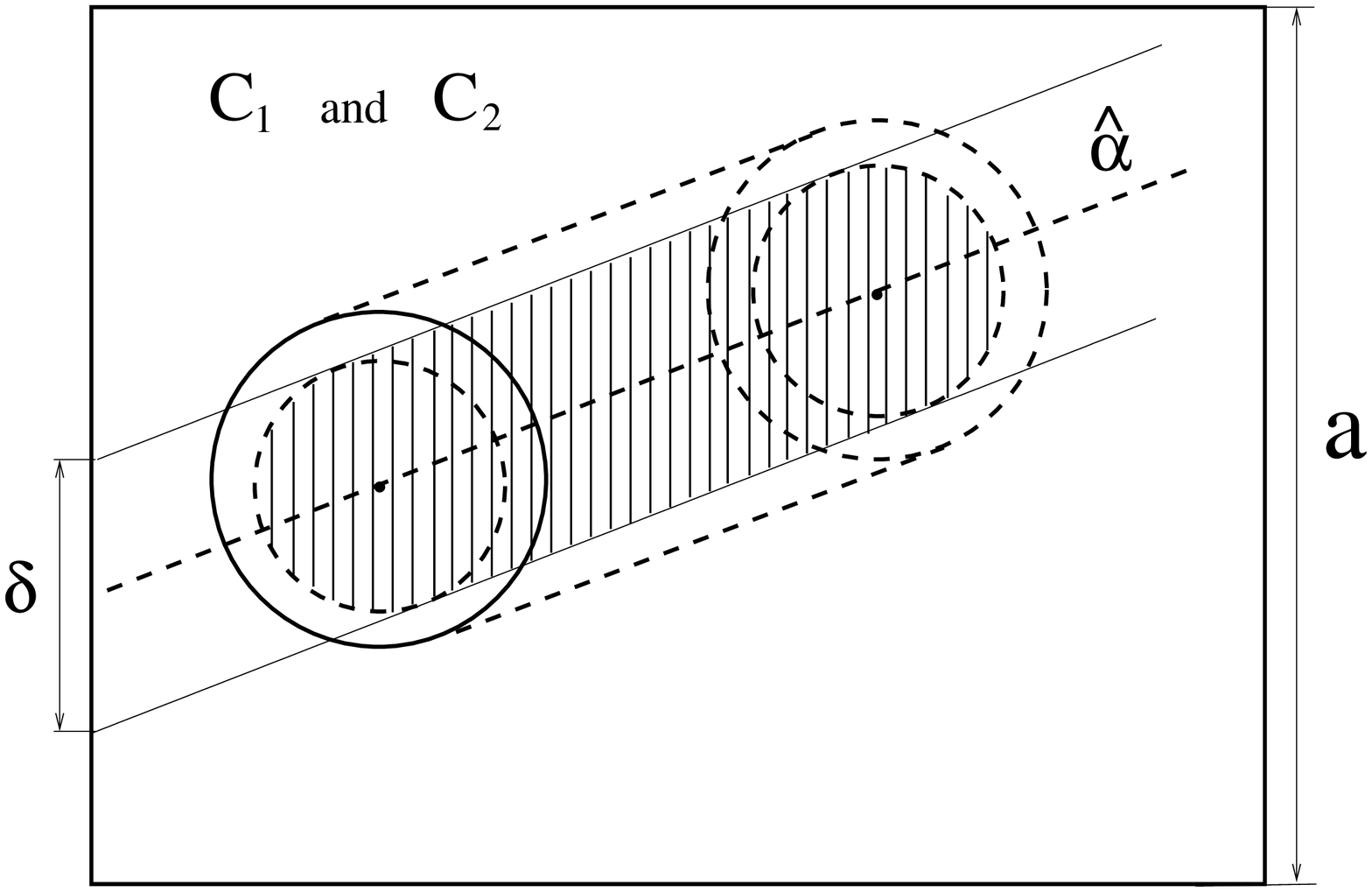,width=14.0cm,height=9cm}
\end{center}
\caption{The intersection of the horizontal planes by the 
vertical plane with irrational direction passing through
the axes of inclined cylinders in Tsarev example. 
}
\label{tsarex2}
\end{figure}

 We will assume that both types of the cylinders have the 
identical forms and two cylinders shown at Fig. \ref{tsarex1}
have the same vertical symmetry plane ${\Pi}$ 
which intersects the horizontal layers in some irrational
direction ${\hat \alpha}$ (Fig. \ref{tsarex2}).

 We can assume now that the dark region on Fig. \ref{tsarex2}
represents actually the "jumping place" where the intersection
curves $\Pi^{\prime} \cap S_{F}$ "jump" from one horizontal layer 
to another (Fig. \ref{tsarex1}) for any plane $\Pi^{\prime}$
parallel to $\Pi$. Let us put now 
$\Pi = \Pi({\bf B})$ and consider the intersections 
$\Pi \cap S_{F}$ as the electron trajectories in 
${\bf p}$-space for given ${\bf B}$.

 We will put for simplicity that the horizontal 
periods of Fermi surface
are orthogonal to each other and we can divide the horizontal 
layers into identical periodic strips parallel to one of the 
periods (say vertical strips formed by the repeated domains shown
at Fig. \ref{tsarex2}). Easy to see that for our specific 
direction ${\hat \alpha}$ any trajectory will either pass
through the strip being at the same layer in ${\mathbb R}^{3}$
or meet the pair of cylinders (of identical form) and jump by 
two layers preserving the layer typer ("positive" or "negative").
The horizontal projection of corresponding orbits parts are
always the same in this situation while the vertical displacements 
can be either zero or the vertical period of $S_{F}$ if the
vertical jump takes place in the corresponding strip. 
The probability of jump is proportional to the ratio $\delta/a$
(Fig. \ref{tsarex2}) and all the non-singular trajectories have 
then the same asymptotic directions.

 It can be also easily seen that the trajectories belonging to
different types of layers go and "jump" in the opposite 
directions being divided by the singular 
trajectories passing through 
the saddle-points on the sides of the cylinders $C_{1}$ and 
$C_{2}$.

 It can be proved, however, that there exist the special values of
${\hat \alpha}$ such that the trajectories will not belong 
here to the straight strips of any finite width as follows from the
numbers theory. In this situation we can not use anymore
the same arguments which we used for calculating of 
$\sigma^{1i}$ and $\sigma^{1i}$ in the case of topologically
regular trajectories and so we don't have formulas
(\ref{sigma11}) and (\ref{sigmai1}) in this situation. 
However, we can still use the fact

$$\langle v^{x}_{gr} \rangle = 0 \,\,\, , \,\,\,
\langle v^{y}_{gr} \rangle \neq 0 \,\,\, , \,\,\,
\langle v^{z}_{gr} \rangle \neq 0 $$
(in the appropriate coordinate system)
also in this case. Using the same arguments for the formula
(\ref{condavf}) we can prove then the following formulae for
the behavior of $\sigma^{ik}({\bf B})$:

\begin{equation}
\label{sigtsar}
\sigma^{ik}({\bf B}) \simeq {n e^{2} \tau \over m^{*}} \,
\left( \begin{array}{ccc}
o(1) & o(1) & o(1) \cr
o(1) & * & * \cr
o(1) & * & *
\end{array} \right)
\end{equation}
which replaces the formula (\ref{sigoptr}) for the case of
Tsarev's ergodic trajectories. Let us omit here all the
details of Tsarev's chaotic trajectories and just point out that 
the asymptotic direction of ergodic trajectory of Tsarev type
can be also observed experimentally due to the same reasons
as in the case of topologically regular trajectories.
However, unlike the topologically regular case, the ergodic
trajectories of Tsarev type are unstable with respect to generic 
small rotations of ${\bf B}$ and do not correspond to 
"stability zones" on the unit sphere. 

 Let us say now some words about more general ergodic 
trajectories of Dynnikov type which do not have any asymptotic
direction in ${\mathbb R}^{3}$. We will not describe here
the corresponding construction (see \cite{dynn4}) and just 
give the main features of such trajectories. 

 First of all,
these trajectories can arise only in the case of magnetic
field of irrationality 3 and the corresponding carriers
have then the genus $\geq 3$. This kind of trajectories are
completely unstable with respect to the small rotations of 
${\bf B}$ and can be observed just for special fixed directions
of ${\bf B}$ in the case of rather complicated Fermi surfaces.
The approximate form of trajectories of this kind is shown
at Fig. \ref{regandch}, b.
The carrier of such trajectory can then resemble the Fermi
surface shown at the Fig. \ref{rank}, d and is homologous to 
zero in the 3-dimensional torus ${\mathbb T}^{3}$. Moreover,
if the genus of the Fermi surface is not very high ($< 6$)
it can always be stated that the corresponding carrier of
open ergodic trajectories is invariant under the involution
${\bf p} \rightarrow - {\bf p}$ (after the appropriate choice
of the initial point in ${\mathbb T}^{3}$). The ergodicity 
of the open trajectories on the carrier gives then immediately
the relations:

$$\langle v^{x}_{gr} \rangle = 0 \,\,\, , \,\,\,
\langle v^{y}_{gr} \rangle = 0 \,\,\, , \,\,\,
\langle v^{z}_{gr} \rangle = 0 $$
for all three components of the group velocity on any of
such trajectories. This important fact leads to the rather 
non-trivial behavior of corresponding contribution to
the conductivity for $B \rightarrow \infty$. Namely, using the
same formula (\ref{sigmaik}) in this case we can show that all
the components of the corresponding contribution to 
$\sigma^{ik}({\bf B})$ become actually zero in the limit
$B \rightarrow \infty$ (\cite{malts}). We can write then for 
this contribution:

\begin{equation}
\label{sigdyn}
\sigma^{ik}({\bf B}) \simeq {n e^{2} \tau \over m^{*}} \,
\left( \begin{array}{ccc}
o(1) & o(1) & o(1) \cr
o(1) & o(1) & o(1) \cr
o(1) & o(1) & o(1)
\end{array} \right)
\end{equation}
for $B \rightarrow \infty$.\footnote{Actually the component
$\sigma^{zz}({\bf B})$ contains the non-vanishing term of order
of $T^{2}/\epsilon_{F}^{2}$ for $B \rightarrow \infty$ for 
non-zero temperatures (\cite{malts}). However, this parameter
is very small for the normal metals and we don't take it here
in the account.}

 We see then that the chaotic trajectories of Dynnikov type
do not give any contribution even for conductivity along
the magnetic field ${\bf B}$ for rather big values of $B$.
In the work \cite{malts} also the special "scaling" asymptotic
behavior of $\sigma^{ik}({\bf B})$ were suggested.
Let us note, however, that the full conductivity tensor
include also the contribution of compact (closed) electron 
trajectories
having the form (\ref{sigcltr}) which presents in general
as the additional contribution in all the cases described above.
We can so claim that the ergodic behavior of Dynnikov type
does not actually completely removes the conductivity along
the magnetic field ${\bf B}$ because of the contribution of
compact trajectories. However, the sharp local minimum in this
conductivity can still be observed in this case since a part
of the Fermi surface will be effectively excluded from the
conductivity in this situation.

 It can be proved (see \cite{dynn7}) that for generic Fermi 
surfaces the measure of directions of magnetic field ${\bf B}$
where the chaotic behavior of Dynnikov type can be found
on the Fermi surface is zero. Let us say that the restriction
on just one energy level is connected closely with the situation
in metals where only the energy levels close to $\epsilon_{F}$
are important. Let us formulate also the more general conjecture
of S.P.Novikov about the total set of "chaotic directions"
on the unit sphere, where we do not restrict the consideration
to just one energy level:

\vspace{0.5cm}
  
 {\bf Conjecture.} (S.P.Novikov).

 The total set of the directions of ${\bf B}$ corresponding
to the chaotic behavior has the measure $0$ for the whole
generic dispersion relation and the Hausdorf dimension strictly
less than $2$.

\vspace{0.5cm}

 Let us mention also that in the paper \cite{dynmal} the 
possibilities of the investigation of total topological 
characteristics of whole dispersion relation $\epsilon({\bf p})$
were discussed. In particular the behavior of electrons
injected in the empty band of semiconductor in the presence of
magnetic field was considered. However, the corresponding
magnetic fields should be extremely high $(\sim 10^{2} Tl)$
in this case which make such experiments very difficult.
Most probably this situation should be considered now
just as theoretical possibility.

\section{The different regimes of conductivity behavior.
Classification.}

 We are going to give now the full classification of possible
conductivity regimes corresponding to different topological types
of trajectories given by system (\ref{dynsyst}). First we will 
need to add here the non-generic cases of irrationality 
1 and 2. The Theorem $3$
should be slightly modified in this case but has the same main
features as in the case of fully irrational magnetic field
(\cite{dynn7}).
Namely, the set of carriers of open trajectories
${\bar S}_{\epsilon}$ can
contain now the two-dimensional tori $T^{2\prime}_{s}$ having
the zero homology class in $H_{2}(T^{3})$ in addition to the
family of parallel tori with non-zero homology classes described  
above. The corresponding covering of these components in $R^{3}$
are "warped" periodic cylinders and all the open
trajectories belonging to these components are purely periodic.
As it is easy to see these components of ${\bar S}_{\epsilon}$
are stable with respect to the small rotations
of ${\bf B}$ in the plane orthogonal to the axis of cylinder   
and disappear after any other small rotation. The part (1)
of the Theorem 2 will be true also for rational or "partly
rational" directions of ${\bf B}$ with some connected energy
interval
$\epsilon_{1}^{\prime}({\bf B}) \leq \epsilon \leq
\epsilon_{2}^{\prime}({\bf B})$. However, the boundary values 
$\epsilon_{1}^{\prime}({\bf B})$, $\epsilon_{2}^{\prime}({\bf B})$
do not necessarily coincide in this case with the values of
piecewise smooth
functions $\epsilon_{1}({\bf B})$, $\epsilon_{2}({\bf B})$ defined
everywhere on $S^{2}$ according to Theorem $3$ (\cite{dynn7}).
Namely, we will have instead the relations
$\epsilon_{1}^{\prime}({\bf B}) \leq \epsilon_{1}({\bf B}) \leq
\epsilon_{2}({\bf B}) \leq \epsilon_{2}^{\prime}({\bf B})$
for all such directions of ${\bf B}$ where all the components
of ${\bar S}_{\epsilon}$ belonging to intervals
$[\epsilon_{1}^{\prime}({\bf B}), \epsilon_{1}({\bf B}))$,
$(\epsilon_{2}({\bf B}), \epsilon_{2}^{\prime}({\bf B})]$
consist of the tori homologous to
zero in $T^{3}$. As we will see this cases can be observed
experimentally for those ${\bf B}$ where the
Fermi level lies in the one of such intervals and only the  
"partly stable" non-compact trajectories exist on the Fermi surface.

 The "partly-stable" cylinders
described above do not intersect the "absolutely stable"
components of ${\bar S}_{\epsilon}$ and all the open
trajectories will still have the same mean direction
if ${\bf B}$ is not orthogonal to corresponding integral plane
$\Gamma_{\alpha}$ (let us mention that all the trajectories lying
on "regular" parallel integral planes in
${\bar S}_{\epsilon} \subset R^{3}$ will be also
periodic with the same period in this case).
The form of the conductivity tensor will still be described by
the formula (\ref{sigoptr})
but the numerical values of dimensionless coefficients will jump
for those non-generic directions of ${\bf B}$ where the situation
described appears. In particular, the asymptotic behavior
(\ref{sigoptr}) will arise on the one-dimensional curves for
those directions of ${\bf B}$ where
$\epsilon_{F} \in
[\epsilon_{1}^{\prime}({\bf B}), \epsilon_{1}({\bf B})) \cup
(\epsilon_{2}({\bf B}), \epsilon_{2}^{\prime}({\bf B})]$
being stable only for rotations of ${\bf B}$ in the corresponding
direction. As follows from the statements above
the corresponding directions of ${\bf B}$ can have at most
the irrationality $2$ and the corresponding one-dimensional
curves are always the parts of the circles orthogonal to some  
integer vector in the reciprocal lattice $\Gamma^{*}$.

 Let us make now a special remark about the "Special directions"
of ${\bf B}$ orthogonal to the integral planes $\Gamma_{\alpha}$
if this direction belongs to the corresponding stability
zone $\Omega_{\alpha}$. The direction of ${\bf B}$ is then
purely rational and all the corresponding open
orbits (if they exist) should be periodic in $R^{3}$. However,
the mean directions of these open orbits can be different
in this case for the different planes orthogonal to the
magnetic field and the corresponding contributions to
conductivity can not be then written in the form
(\ref{sigoptr}) in the same coordinate system. The
conductivity tensor $\sigma^{ik}$ can have then the full rank
in the limit $B\tau \rightarrow \infty$ and the conductivity  
remains constant for all the directions in $R^{3}$ in the
strong magnetic field limit. This situation, however, is
completely unstable and disappear after any small rotation
of ${\bf B}$.

\begin{figure}
\begin{center}
\epsfig{file=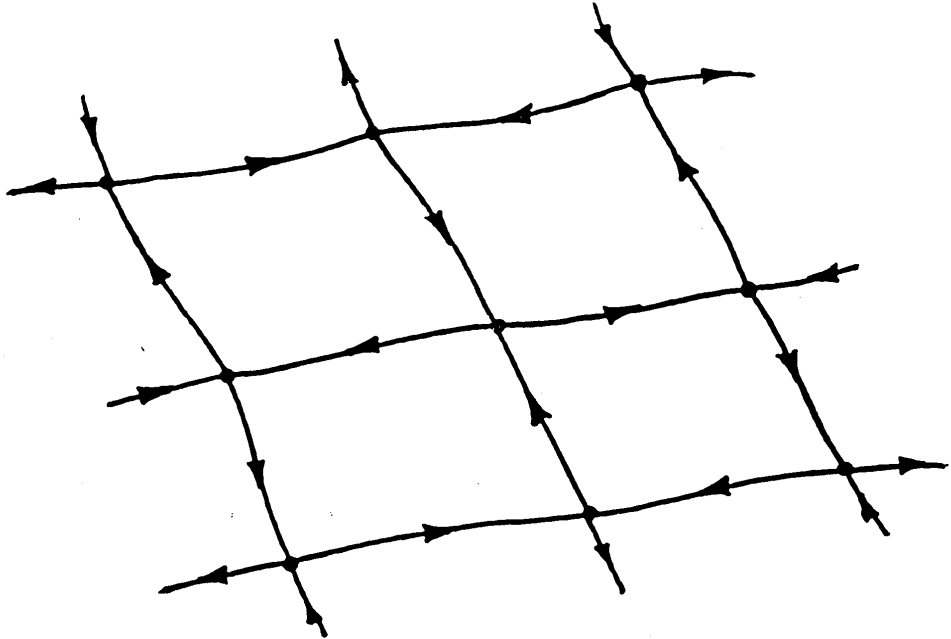,width=14.0cm,height=10cm}
\end{center}
\caption{The "singular periodic net" of open trajectories
in the plane orthogonal to ${\bf B}$.
}
\label{singnet}
\end{figure}

 The second possibility in the case of such directions is that
all the open trajectories
become singular and form the "singular periodic nets" in the
planes orthogonal to ${\bf B}$ (see Fig. \ref{singnet}). 
The asymptotic
behavior of conductivity is described then by formula
(\ref{sigcltr}) but is also completely unstable and changes
to (\ref{sigoptr}) after any small rotation of ${\bf B}$. 

 Let us add that the same situations can arise also
on the stable two-dimensional tori $T^{2}_{i}$ in this case
where the directions of open orbits can not be defined anymore
as the intersection of $\Gamma_{\alpha}$ with the plane
$\Pi({\bf B})$. Such, we can have either the "singular nets"
or the regular periodic open orbits on these tori for this   
special direction. Also we can have the open
orbits with different mean directions on different tori  
in this case but the number of such tori should be $\geq 4$
for any physical type of dispersion relations. This situation
can thus also be observed only for rather complicated Fermi
surfaces.
 
\begin{figure}
\begin{center}
\epsfig{file=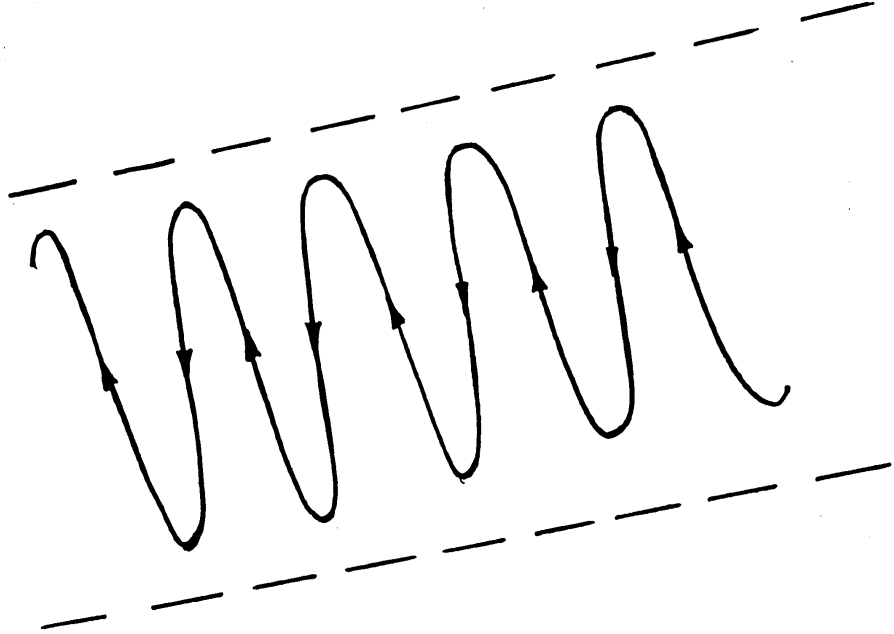,width=14.0cm,height=10cm}
\end{center}
\caption{The wide finite strip in the plane $\Pi({\bf B})$
containing the open orbits for ${\bf B}$ close to
"Special rational direction" within the stability zone.
}
\label{widestrip}
\end{figure}

 Let us mention also that for the directions of ${\bf B}$
close to this special rational directions the widths of the 
straight strips containing the regular open orbit can become
very big in the planes $\Pi({\bf B})$ (see Fig. \ref{widestrip}). 

 So from
physical point of view the conductivity phenomena will not 
"feel" the mean directions of open orbits for the directions
of ${\bf B}$ close enough to these special points
even for rather big (but finite) values of magnetic field.
Instead, the oscillations of the trajectory within the strip
(Fig. \ref{widestrip}) will be essential for conductivity up 
to the rather
big values of $B$ such that $\omega_{B}\tau \sim L/p_{0}$ 
(where $L$ is the width of the strip and $p_{0}$ is the size
of the Brillouin zone). These situation exists, however, if
the open orbits with corresponding direction exist also for
${\bf B} = {\bf B}_{0}$ where ${\bf B}_{0}$ is the special
rational direction in the stability zone. For ${\bf B}$
close to ${\bf B}_{0}$ we will observe then exactly this
direction up to the values of $B$ such that
$\omega_{B}\tau \sim L/p_{0}$ and then the regime will change 
to the common situation corresponding to given stability zone.
Experimentally we will observe then the "small spots" around
these directions on the unit sphere where the anisotropy of  
$\sigma^{ik}$ corresponds to the direction of orbits for
${\bf B} = {\bf B}_{0}$. In the most complicated case when we
have the open orbits with different mean directions for
${\bf B} = {\bf B}_{0}$ we will have then the finite conductivity
for all the directions in $R^{3}$ in the corresponding spot
for rather big values of $B$.
 
  For the "special rational directions" corresponding to the
"singular net" on the stable tori $T^{2}_{i}$ the behavior  
of $\sigma^{ik}$ will correspond to the common form for
a given stability zone even for ${\bf B}$ very close to
${\bf B}_{0}$. However, the measure of open orbits will    
tend to zero as ${\bf B} \rightarrow {\bf B}_{0}$ in the
stability zone. The dimensionless coefficients ($*$) in the
formula (\ref{sigoptr}) will vanish then for
${\bf B} \rightarrow {\bf B}_{0}$ although the integral plane
$\Gamma_{\alpha}$ will be observable up to
${\bf B} = {\bf B}_{0}$.

  We mention here at last that both cases when
${\bf B}^{\alpha}_{0}$
belongs or does not belong to the corresponding stability zone
$\Omega_{\alpha}$ are possible in the examples.

 We can describe now the total picture for the angle diagram
of conductivity in normal metal in the case of geometric
strong magnetic field limit.
Namely, we can observe the following objects on the
unit sphere parameterizing the directions of ${\bf B}$:

\vspace{0.5cm}

 1) The "stability zones" $\Omega_{\alpha}$ corresponding to
some integral planes $\Gamma_{\alpha}$ in the reciprocal lattice
("Topological Quantum numbers"). This "Topological Type" of open
trajectories
is stable with respect to small rotations of ${\bf B}$ and this
is the only open orbits regime which can have a non-zero measure
on the unit sphere. All the "stability zones" have the piecewise
smooth boundaries on $S^{2}$ and are given by the condition

$$\epsilon_{1}({\bf B}) \,\,\, \leq \,\,\, \epsilon_{F} \,\,\,
\leq \,\,\, \epsilon_{2}({\bf B})$$
where $\epsilon_{1}({\bf B})$, $\epsilon_{2}({\bf B})$ are
the piecewise smooth functions defined in the Theorem $3$.
Generally speaking, this set is not everywhere dense
anymore being just a subset of the corresponding
set for the whole dispersion relation and have some
rather complicated geometry on the unit sphere.

 The corresponding behavior of conductivity
is described by the formula (\ref{sigoptr}) and reveals the strong
anisotropy in the planes orthogonal to the magnetic field.  
For rather complicated Fermi surfaces we can observe also the
"sub-boundaries" of the stability zones where the coefficients
in (\ref{sigoptr}) have the sharp "jump".

\vspace{0.5cm}

 2) The net of the one-dimensional curves containing directions
of irrationality $\leq 2$ where the additional two-dimensional
tori (homologous to zero in $T^{3}$) can appear. The corresponding
parts of the net are always the parts of the big (passing
through the center of $S^{2}$) circles orthogonal to some
reciprocal lattice vector where the condition

$$\epsilon_{1}^{\prime}({\bf B}) \,\,\, \leq \,\,\, \epsilon_{F}
\,\,\, \leq \,\,\, \epsilon_{2}^{\prime}({\bf B})$$
is satisfied. The asymptotic behavior of conductivity is given
again by the formula (\ref{sigoptr}).

 Let note also that these special curves on $S^{2}$ can be
considered actually as the reminiscent of the bigger stability
zones if we don't restrict ourselves just by one Fermi surface.
The structure of such sets thus can be used to get more information
about the corresponding total structure for the whole dispersion
relation in metal. In particular, the mean direction of such open
orbits coincides with the mean direction of the generic open
orbits in the intersections of the net with the stability
zones (except the "Special rational directions"). The corresponding
conductivity tensor is given then by the same formula (\ref{sigoptr})
where the dimensionless coefficients "jump" on the curves of the
net.

\vspace{0.5cm}

 3) The "Special rational directions".

 Let us remind that we call the special
rational direction the direction of ${\bf B}$
orthogonal to some plane $\Gamma_{\alpha}$ in case when this
direction belongs to the same stability
zone on the unit sphere. We can have here
all the possibilities described earlier for this situation
(i.e. regular behavior with vanishing coefficients in
(\ref{sigoptr}), spots with isotropic or anisotropic behavior   
of conductivity different from the given by corresponding
"Topological quantum numbers", "partly stable"
isotropic or anisotropic addition to the (\ref{sigoptr}), etc.)

\vspace{0.5cm}

 4) The chaotic open orbits of Tsarev type
(${\bf B}$ of irrationality $2$).

 We can have points on the unit sphere where the open orbits
are chaotic in Tsarev sense. All open trajectories still have
the asymptotic direction in this case and the conductivity
reveals the strong anisotropy in the plane orthogonal to
${\bf B}$ as $B \rightarrow \infty$. The $B$ dependence, however
is slightly different from the formula (\ref{sigoptr}) in   
this case.

\vspace{0.5cm}

 5) The chaotic open orbits of Dynnikov type
(${\bf B}$ of irrationality $3$).

 For some points on $S^{2}$ we can have the chaotic open
orbits of Dynnikov type on the Fermi surface.
At these points the local
minimum of conductivity along the magnetic field is expected.
The conductivity along ${\bf B}$ however remains finite as
$B \rightarrow \infty$ in general situation because of the
contribution of compact trajectories. 

\vspace{0.5cm}

 6) At last we can have the open regions on the unit sphere
where only the compact trajectories on the Fermi level are
present. The asymptotic behavior of conductivity tensor is
given then by the formula (\ref{sigcltr}).

\vspace{0.5cm}

 Let now point out some new features connected with the
"magnetic breakdown" (self-intersecting Fermi surfaces) which   
can be observed for rather strong magnetic fields. Up to this
point it has been assumed throughout that different parts
of the Fermi surface do not intersect with each other.
However, it is possible for some special lattices that the
different components of the Fermi surface (parts corresponding
to different conductivity bands) come very close to each other
and may have an effective "reconstruction" as a result of
the "magnetic breakdown" in strong magnetic field limit.
In this case we can have the situation of the electron
motion on the self-intersecting Fermi surface such that
the intersections with other pieces do not affect at all the 
motion on one component. (The physical conditions for the 
corresponding values of $B$ can be found in \cite{etm}).  
In this case the picture described above should be considered   
independently for all the non-selfintersecting pieces    
of Fermi surface and we can have simultaneously several 
independent angle diagrams of this form on the unit sphere.
Such we can have here the overlapping stability zones
where the open orbits can have different mean directions.
The correspondent conductivity tensor will then be given
just as a sum of all conductivity tensors corresponding
to different non-selfintersecting components.
(The problem of the magnetic breakdown was brought to
the authors' attention by M.I.Kaganov.)


\begin{thebibliography}{99}

\bibitem{lifazkag} I.M.Lifshitz, M.Ya.Azbel, M.I.Kaganov.
{\it Sov. Phys. JETP} {\bf 4}, 41 (1957).

\bibitem{lifpes1} I.M.Lifshitz, V.G.Peschansky.
{\it Sov. Phys. JETP} {\bf 8}, 875 (1959).

\bibitem{lifpes2} I.M.Lifshitz, V.G.Peschansky.
{\it Sov. Phys. JETP} {\bf 11}, 137 (1960).

\bibitem{ag1} N.E.Alexeevsky, Yu.P.Gaidukov.
{\it Sov. Phys. JETP} {\bf 8}, 383 (1959).

\bibitem{ag2} N.E.Alexeevsky, Yu.P.Gaidukov.
{\it Sov. Phys. JETP} {\bf 9}, 311 (1959).

\bibitem{ag3} N.E.Alexeevsky, Yu.P.Gaidukov.
{\it Sov. Phys. JETP} {\bf 10}, 481 (1960).

\bibitem{gaid} Yu.P.Gaidukov.
{\it Sov. Phys. JETP} {\bf 10}, 913 (1960).

\bibitem{lifkag1} I.M.Lifshitz, M.I.Kaganov.
{\it Sov. Phys. Usp.} {\bf 2}, 831 (1960).

\bibitem{lifkag2} I.M.Lifshitz, M.I.Kaganov.
{\it Sov. Phys. Usp.} {\bf 5}, 411 (1962).

\bibitem{etm} I.M.Lifshitz, M.Ya.Azbel, M.I.Kaganov.
Electron Theory of Metals. Moscow, Nauka, 1971.
Translated: New York: Consultants Bureau, 1973.

\bibitem{abr} A.A.Abrikosov.
Fundamentals of the Theory of Metals.
"Nauka", Moscow (1987). Translated:
Amsterdam: North-Holland, 1998.

\bibitem{nov1} S.P.Novikov.
{\it Russian Math. Surveys} {\bf 37}, 1 (1982).

\bibitem{nov2} S.P.Novikov.
Proc. Steklov Inst. Math. 1 (1986).

\bibitem{nov3} S.P.Novikov. "Quasiperiodic structures in topology".
Proc. Conference "Topological Methods in Mathematics",
dedicated to the 60th birthday of J.Milnor, June 15-22, S.U.N.Y.
Stony Brook, 1991. Publish of Perish, Houston, TX, pp. 223-233 (1993).

\bibitem{nov4} S.P.Novikov. Proc. Conf. of Geometry,
December 15-26, 1993, Tel Aviv University (1995).

\bibitem{zorich} A.V.Zorich.
{\it Russian Math. Surveys} {\bf 39}, 287 (1984). 

\bibitem{dynn1} I.A.Dynnikov.
{\it Russian Math. Surveys} {\bf 57}, 172 (1992).

\bibitem{dynn2} I.A.Dynnikov.
{\it Russian Math. Surveys} {\bf 58} (1993).

\bibitem{dynn3} I.A.Dynnikov. "A proof of Novikov's conjecture
on semiclassical motion of electron."
{\it Math. Notes} {\bf 53}:5, 495 (1993).

\bibitem{novmal1} S.P.Novikov, A.Ya.Maltsev.
{\it ZhETP Lett.} {\bf 63}, 855 (1996).

\bibitem{dynn5} I.A.Dynnikov.
"Surfaces in 3-Torus: Geometry of plane sections."  
Proc.of ECM2, BuDA, 1996.

\bibitem{dynn4} I.A.Dynnikov.
"Semiclassical motion of the electron. A proof of the Novikov
conjecture in general position and counterexamples."
American Mathematical Society Translations, Series 2, Vol. 179,
Advances in the Mathematical Sciences. Solitons, Geometry, and
Topology: On the Crossroad. Editors: V.M.Buchstaber,
S.P.Novikov. (1997)

\bibitem{dynmal} I.A.Dynnikov, A.Ya.Maltsev.
{\it ZhETP} {\bf 85}, 205 (1997).

\bibitem{malts} A.Ya. Maltsev.
{\it ZhETP} {\bf 85}, 934 (1997).

\bibitem{novmal2} S.P.Novikov, A.Ya.Maltsev.
{\it Physics-Uspekhi} {\bf 41}(3), 231 (1998).

\bibitem{zorich2} A.V.Zorich.
Proc. "Geometric Study of Foliations" (Tokyo, November 1993)/
ed. T.Mizutani et al. Singapore: World Scientific, 479-498
(1994).

\bibitem{dynn7} I.A.Dynnikov.
{\it Russian Math. Surveys} {\bf 54}, 21 (1999).

\bibitem{tsarev} S.P.Tsarev.
Private communication. (1992-93).

\bibitem{nov5} S.P.Novikov. 
{\it Russian Math. Surveys} {\bf 54}:3, 1031 (1999).

\bibitem{DeLeo} R.D.Leo. PhD Theses. University of Maryland.
Department of Math., College Park, MD 20742, USA.

\bibitem{malnov3} A.Ya.Maltsev, S.P.Novikov,
ArXiv: math-ph/0301033,
to appear in Special Volume of Bulletin of Braz. Math. Society.

\bibitem{malts2}  A.Ya. Maltsev,
Arxiv: cond-mat/0302014 

\bibitem{arnold} V.I.Arnold. {\it Functional analysis and
its appilcations} {bf 25}:2 (1991).

\bibitem{sinkhan} Ya.G.Sinai, K.M.Khanin.
{\it Functional analysis and its appilcations}
{bf 26}:3 (1992).

\bibitem{pippard} A.B.Pippard. Phil. Trans. Roy. Soc.,
{\bf A250}, 325 (1957).

\bibitem{cohfal} M.H.Cohen, L.M.Falicov. Phys. Rev. Lett.
{\bf 7}, 231 (1961).


\end{thebibliography}
\end{document}